\documentclass[journal,12pt,onecolumn,draftclsnofoot,]{IEEEtran}
\usepackage{graphicx}

\usepackage[utf8]{inputenc}
\usepackage[T1]{fontenc}
\usepackage{cite}
\usepackage{graphicx}
\usepackage{psfrag}
\usepackage{subfigure}
\usepackage[caption=false]{subfig}
\usepackage{url}
\usepackage{ifthen}
\usepackage[cmex10]{amsmath}
\usepackage{array}
\usepackage{amssymb}
\usepackage{amsthm}
\usepackage{mathtools}
\usepackage{amsfonts}
\usepackage{graphicx}
\usepackage{epstopdf}
\usepackage{algorithm}
\usepackage{algorithmic}
\newtheorem{proposition}{Proposition}
\newtheorem{lemma}{Lemma}
\newtheorem{assumption}{Assumption}

\newtheorem{definition}{Definition}
\newtheorem{theorem}{Theorem}

\usepackage{setspace}
\usepackage{amscd}
\usepackage{mathrsfs}
\usepackage{epsfig}
\usepackage{color}
\usepackage{textcomp}
\usepackage{comment}
\usepackage{xcolor}
\usepackage{pgf}
\usepackage{pgfplots}
\pgfplotsset{compat=1.9}
\usepackage{tikz}
\usetikzlibrary{decorations.pathreplacing}
\usetikzlibrary{decorations.markings}
\usetikzlibrary{patterns}
\usepackage[outline]{contour}
\contourlength{1.2pt}
\pgfplotsset{compat=1.18} 
\usetikzlibrary{plotmarks}
\usetikzlibrary{spy}
\usetikzlibrary{calc}
\usetikzlibrary{shapes,plotmarks,positioning,fit,chains,tikzmark,arrows,hobby,arrows.meta}
\usetikzlibrary{decorations.pathreplacing,calligraphy}
\usepgfplotslibrary{groupplots}

\usepackage[font=scriptsize]{caption}

\usepackage{import}

\usepackage[toc, notranslate, acronym]{glossaries}
\usepackage{glossaries-extra}
\setabbreviationstyle[acronym]{long-short}
\newacronym{4g}{4G}{fourth generation}
\newacronym{5g}{5G}{fifth generation}
\newacronym{6g}{6G}{sixth generation}
\newacronym{iid}{i.i.d.}{independent and identically distributed}
\newacronym{llm}{LLM}{large language model}
\newacronym{kl}{KL}{Kullback Leibler}
\newacronym{occ}{OCC}{online conformal compression}
\newacronym{cp}{CP}{conformal prediction}
\newacronym{ocp}{OCP}{online conformal prediction}
\newacronym{ai}{AI}{artificial intelligence}
\newacronym{bcc}{BCC}{block conformal compression}

\newacronym{ocrdc}{OCRDC}{online conformal rate-distortion compression}
\newacronym{ocsc}{OCSC}{online conformal sparse compression}

\DeclareMathOperator*{\argmax}{argmax}
\DeclareMathOperator*{\argmin}{argmin}

\usepackage{dsfont}
\usepackage{bbm}

\usepackage{graphicx}
\usepackage{subcaption}

\title{Prediction-Powered Communication with Distortion Guarantees}
\author{ Matteo Zecchin, Unnikrishnan Kunnath Ganesan, Giuseppe Durisi, Petar Popovski, and Osvaldo Simeone
\thanks{Matteo Zecchin and Osvaldo Simeone are with the  King's Communications, Learning \& Information Processing (KCLIP) lab within the Centre for Intelligent Information Processing Systems (CIIPS), Department of Engineering, King's College London, London WC2R 2LS, U.K. (e-mail: \{matteo.1.zecchin, osvaldo.simeone\}@kcl.ac.uk). Unnikrishnan Kunnath Ganesan and Giuseppe Durisi are with the Department of Electrical Engineering, Chalmers University of Technology, 41296 Gothenburg, Sweden.  (e-mail: \{kunnathg, durisi\}@chalmers.se). Petar Popovski is with the Department of Electronic Systems, Aalborg University, Aalborg, Denmark (e-mail: petarp@es.aau.dk). \\ The work of  M. Zecchin and O. Simeone was  supported by the Open Fellowships of the EPSRC (EP/W024101/1) and by the EPSRC project EP/X011852/1.}}
\begin{document}
\maketitle

\begin{abstract}

 The development of 6G wireless systems is taking place alongside the development of increasingly intelligent wireless devices and network nodes. The changing technological landscape is motivating a rethinking of classical Shannon information theory that emphasizes semantic and task-oriented paradigms. In this paper, we study a prediction-powered communication setting, in which devices, equipped with artificial intelligence (AI)-based predictors, communicate under zero-delay constraints with  strict distortion guarantees. Two classes of distortion measures are considered: (i) outage-based metrics, suitable for tasks tolerating occasional packet losses, such as real-time control or monitoring; and (ii) bounded distortion metrics, relevant to semantic-rich tasks like text or video transmission. We propose two zero-delay compression algorithms leveraging online conformal prediction to provide per-sequence guarantees on the distortion of reconstructed sequences over error-free and packet-erasure channels with feedback. For erasure channels, we introduce a doubly-adaptive conformal update to compensate for channel-induced errors and derive sufficient conditions on erasure statistics to ensure distortion constraints. Experiments on semantic text compression validate the approach, showing significant bit rate reductions while strictly meeting distortion guarantees compared to state-of-the-art prediction-powered compression methods.
\end{abstract}

\section{Introduction}
\subsection{Context and Motivation}

A central feature of various 6G visions is connected intelligence, that is, wireless connectivity among artificial intelligence (AI)-powered wireless devices and network nodes. The development of solutions for this new communication setting shifts the design  objective away from bit-perfect replication to the exchange of semantically meaningful and task-relevant information \cite{shi2023task,shen2024large}. Instead of indiscriminately sending the entire raw data stream, distributed agents equipped with AI models and contextual knowledge use 6G connectivity to exchange features that are essential for semantic understanding or for successful task execution \cite{popovski2020semantic,strinati2024goal,guo2024distributed}.
Consequently, next-generation communication protocols are increasingly envisioned to be designed according to distortion measures that capture application-level requirements, rather than traditional symbol- or bit-level fidelity. Such measures emphasize the preservation of semantic content in multimedia information such as  text or video~\cite{getu2023making}.

The emergence of connected intelligence requires a revision of the traditional communication model to include the operation of AI modules at the communicating parties. Concretely, the task-specific fidelity requirements may translate into different types of distortion constraints, including:  (\emph{i}) \emph{outage-based metrics}, which allow for a limited fraction of losses and are well-suited for tasks such as real-time control; and  (\emph{ii}) \emph{general bounded distortion metrics}, which are necessary for tasks requiring the faithful transmission of semantically rich data, such as language or visual content.

Beyond semantic fidelity, task-oriented communication must also satisfy stringent communication-centric constraints, notably on  latency and reliability. For instance, safety-critical applications like autonomous driving, industrial automation, and collaborative robotics demand that reliability constraints be met under strict zero-delay operation, with guarantees that hold for every transmitted sequence \cite{pourkabirian2024vision}. However, current AI-driven semantic communication methods, while powerful at extracting task-relevant features, typically lack mechanisms for providing deterministic, certifiable guarantees. Addressing this gap calls for new compression and transmission schemes that combine the adaptivity of AI-based prediction with rigorous, finite-sample reliability guarantees. Addressing this gap is the goal of this work.

\subsection{Related Work}

Classical rate-distortion theory provides a principled foundation for lossy compression \cite{shannon1959coding,berger2003rate}. This framework covers both settings with full knowledge of the source distribution and universal compression algorithms that operate without prior statistical information \cite{davisson1973universal,ziv2003coding}. Universal schemes can adapt to any stationary, ergodic source and are provably asymptotically optimal, achieving performance that approaches the theoretical rate-distortion limit as the block length grows \cite{jalali2008rate,jalali2011lossy}. A notable subclass is given by \emph{fixed-slope} schemes, which provide the flexibility to operate at different points along the rate-distortion curve \cite{yang1997fixed}. However, these methods are inherently designed for offline, block-based processing, and their guarantees only hold asymptotically for large block lengths. As such, they cannot meet the strict finite-horizon and zero-delay requirements imposed by many real-time applications envisioned for 6G networks.  

To enable real-time operation, online learning methods have been explored in the context of compression, giving rise to zero-delay algorithms \cite{linder2001zero,gyorgy2004efficient,matloub2006universal}. Online learning addresses sequential decision-making in distribution-free environments \cite{shalev2012online}, and its techniques have been leveraged to design adaptive compression schemes with performance guarantees expressed in terms of \emph{regret}. Specifically, regret bounds compare the cumulative distortion of an online compressor to that of the best fixed scheme in hindsight \cite{linder2001zero,gyorgy2004efficient,matloub2006universal}. While such results ensure that online algorithms are asymptotically competitive with their offline counterparts, they do not translate into absolute, per-sequence guarantees. This limitation is critical in safety-sensitive applications, where finite-sample performance certificates are required.  

The framework of \emph{online conformal prediction} has recently emerged as a powerful alternative for achieving such guarantees in sequential decision-making tasks \cite{gibbs2021adaptive,feldman2023achieving}. Unlike classical statistical approaches that rely on distributional assumptions, online conformal methods adaptively calibrate an algorithm's operation based on real-time feedback, ensuring that reliability metrics are satisfied \emph{for every sequence}, even under distribution shifts. Importantly, these methods preserve the asymptotic regret guarantees of online optimization, while introducing the concept of \emph{gradient equilibrium}, which can be used to establish deterministic, finite-sample performance certificates \cite{angelopoulos2024online,angelopoulos2025gradient}. This makes online conformal prediction a natural tool for designing compression schemes that meet the zero-delay and application-specific reliability requirements of task-oriented 6G communication systems.

\subsection{Contributions}
Motivated by the need to complement task-oriented and semantic communication with {deterministic performance guarantees}, we introduce prediction-powered communication, based on zero-delay lossy compression algorithms that leverage online conformal prediction to ensure per-sequence control of distortion under arbitrary bounded metrics. Our main contributions are summarized as follows:  
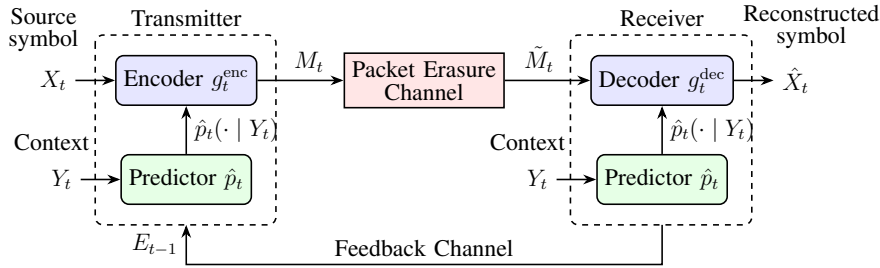
\begin{figure}[!t]
\centering
\scalebox{0.75}{\tikzset{every picture/.style={line width=0.8pt}}
\begin{tikzpicture}[x=1pt,y=1pt,>=Stealth]

\tikzstyle{block} = [rectangle, draw, rounded corners, minimum height=25pt, minimum width=60pt, text centered, fill=blue!10]
\tikzstyle{channel} = [rectangle, draw, minimum height=25pt, minimum width=60pt, text centered, fill=red!10]
\tikzstyle{predictor} = [rectangle, draw, rounded corners, minimum height=25pt, minimum width=60pt, text centered, fill=green!10]
\tikzstyle{arrow} = [->, thick]
\tikzstyle{dashedbox} = [rectangle, draw, dashed, rounded corners, inner sep=5pt]

\node[block] (encoder) at (0,0) {Encoder $g_t^{\rm enc}$};
\node[predictor] (predictorT) at (0,-50) {Predictor $\hat{p}_t$};
\node[left=20pt of encoder] (X) {$X_t$};
\node[left=20pt of predictorT] (C) {$Y_t$};

\draw[dashed, rounded corners] 
    ($(encoder.north west)+(-10,10)$) rectangle ($(predictorT.south east)+(12,-10)$);

\node[channel] (channel) at (120,0) {\shortstack{Packet Erasure \\ Channel}};

\node[block] (decoder) at (240,0) {Decoder $g_t^{\rm dec}$};
\node[predictor] (predictorR) at (240,-50) {Predictor $\hat{p}_t$};
\node[left=20pt of predictorR] (C2) {$Y_t$};
\node[right=20pt of decoder] (Xhat) {$\hat{X}_t$};

\draw[dashed, rounded corners] 
    ($(decoder.north west)+(-10,10)$) rectangle ($(predictorR.south east)+(12,-10)$);

\draw[arrow] (encoder.east) -- (channel.west) node[pos=.6, above] {$M_t$};
\draw[arrow] (channel.east) -- (decoder.west) node[pos=.4, above] {$\tilde{M}_t$};

\draw[arrow] 
    ([yshift=-10pt]predictorR.south) -- ++(0,-20)   
    -- ++(-240,0)        node[midway, above] {Feedback Channel}             
    -- ([yshift=-10pt]predictorT.south)              
    node[midway, left] {$E_{t-1}$};    

\draw[arrow] (X) -- (encoder);
\draw[arrow] (predictorT) -- (encoder)node[midway, right] {$\hat{p}_t(\cdot \mid Y_t)$};
\draw[arrow] (C) -- (predictorT);

\draw[arrow] (predictorR) -- (decoder)node[midway, right] {$\hat{p}_t(\cdot \mid Y_t)$};
\draw[arrow] (C2) -- (predictorR);
\draw[arrow] (decoder) -- (Xhat);

\node[above=10pt of encoder] {Transmitter};
\node[above=10pt of decoder] {Receiver};
\node[above=1pt of X, xshift=-5pt] {\shortstack{Source\\symbol}};
\node[above=1pt of Xhat, xshift=8pt] {\shortstack{Reconstructed\\symbol}};
\node[above=1pt of C, xshift=-5pt] {\shortstack{Context}};
\node[above=1pt of C2, xshift=-5pt] {\shortstack{Context}};
\end{tikzpicture}}
\caption{Zero-delay prediction-based online lossy compression over packet channels: A shared prediction model $\hat{p}_t$ and the current context $Y_t$ are used to encode the current source symbol $X_t$ into a binary message $M_t$, which is transmitted through a packet erasure channel with feedback. The received message $\tilde{M}_t$ is mapped to a reconstructed symbol $\hat{X}_t$ based on the output of the prediction model $\hat{p}_t$ and the context $Y_t$.}
\label{fig:system_diagram}
\end{figure}
\begin{itemize}
\item Focusing on the communication setting illustrated in Fig. \ref{fig:system_diagram}, we propose two zero-delay online lossy compression schemes: \emph{\gls{ocsc}}, tailored for outage-based distortion metrics suitable for tasks that can tolerate limited packet losses, and \emph{\gls{ocrdc}}, designed for general bounded distortion measures relevant to semantic-rich tasks such as text or video transmission.  
\item We prove that, for any source sequence and any sequence of predictors, both schemes guarantee deterministic, finite-sample bounds on the distortion of the reconstructed sequence.  
\item We extend these schemes to operate reliably over erasure channels with feedback by developing channel-adaptive (CA) versions (CA-OCSC and CA-OCRDC). Central to this extension is a novel \emph{doubly-adaptive conformal update rule}, which dynamically compensates for channel-induced errors. We derive rigorous performance bounds for deterministic and stochastic erasure channels, with and without memory, which relate to different types of packet-level models and scheduling protocols in current 5G and future 6G networks.

\item We validate the proposed schemes through extensive experiments on semantic text compression tasks under both outage-based and general distortion metrics. The results demonstrate that our methods not only satisfy the target distortion guarantees but also achieve significant bit rate reductions compared to state-of-the-art prediction-powered compressors, in both error-free and noisy channel conditions.  
\end{itemize}

A preliminary version of this work was presented in the workshop paper \cite{ganesan2025online}, which introduces \gls{ocsc} only, without covering \gls{ocrdc}, CA-OCSC and CA-OCRDC.  The remainder of the paper is organized as follows. In Section~\ref{sec:sys_model}, we introduce the system model, the online lossy compression problem, and the relevant performance metrics. In Sections~\ref{sec:occ_error_free} and~\ref{sec:ConfRD_noiseless}, we detail the \gls{ocsc} and \gls{ocrdc} schemes, respectively, and establish their finite-sample distortion guarantees in error-free communication settings. In Section~\ref{sec:noisy}, we extend these schemes to erasure channels, presenting the channel-adaptive variants and deriving sufficient conditions for meeting distortion guarantees under both deterministic and stochastic channel models. In Section~\ref{sec:exp}, we report experimental evaluations on text compression, benchmarking against existing prediction-powered methods. Finally, in Section~\ref{sec:conclusion}, we conclude the paper and outline directions for future research at the intersection of task-oriented communication, AI-driven prediction, and deterministic reliability in 6G networks.

\section{System Model}
\label{sec:sys_model}

We study the problem of compressing an individual sequence of source symbols $X_1, X_2, \dots$ taking values in a discrete and finite alphabet $\mathcal{X}$, using zero-delay encoding and transmission over a packet erasure channel, i.e., an erasure channel with ACK/NACK feedback. This is used as a model for wireless transmission over low-latency links with small packets, already pursued in 5G and planned to be part of 6G. No probabilistic assumptions are imposed on the source sequence $X_1, X_2, \dots$. 

As illustrated in Fig.~\ref{fig:system_diagram}, at each time $t$, the transmitter observes the symbol $X_t$ and applies a zero-delay, time-dependent encoding function $g_t^{\rm enc}:\mathcal{X}\to\{0,1\}^*$ that maps the symbol $X_t$ to a message $M_t$ of length $b_t$ bits. Here, we use the notation $\{0,1\}^*$ to denote the set of all finite bit strings. To model the transmission of a packet, we assume that the message is conveyed to the receiver over a packet erasure channel. Specifically, let $E_t \in \{0,1\}$ denote the erasure event at time $t$, with $E_t = 1$ representing a packet erasure and $E_t = 0$ representing correct reception. The message received by the decoder is  
\begin{align}
    \label{eq:channel_output}
    \tilde{M}_{t} =
    \begin{cases}
        M_t, & \text{if $E_t = 0$}, \\
        \circledast,   & \text{if $E_t = 1$},
    \end{cases}
\end{align}
where $\circledast$ denotes an erasure message. Unless stated otherwise, we make no probabilistic assumptions on the erasure process $E_1,E_2,\dots$.

Upon reception of the channel output $\tilde{M}_t$ in \eqref{eq:channel_output}, the decoder applies a zero-delay decoding function $g_t^{\rm dec}:\{0,1\}^*\cup\{\circledast\}\to \mathcal{X}$ to obtain an estimate $\hat{X}_t$ of the symbol $X_t$ from message~$\tilde{M}_t$. We assume a noiseless ACK/NACK feedback channel, through which the receiver reliably communicates the erasure variable $E_t$ back to the transmitter. The feedback ensures that at time $t$ both the encoder and the decoder know the entire erasure history $E^{t-1}=(E_1,\dots,E_{t-1})$. 
 
We consider a prediction-powered coding framework, where the encoding and decoding functions $g_t^{\rm enc}$ and $g_t^{\rm dec}$ are designed based on the output of a prediction model for the source sequence. At each time $t$, the predictor $\hat{p}_t$ maps the current context $Y_t$ to a distribution  $\hat{p}_t(\cdot \mid Y_t) \in \Delta(\mathcal{X})$ over the next symbol $X_t$, where $\hat{p}_t(x \mid Y_t)$ denotes the probability assigned to symbol $x \in \mathcal{X}$ and $\Delta(\mathcal{X})$ is the probability simplex for the alphabet $\mathcal{X}$. The context $Y_t$ can include the past symbols $X^{t-1}=(X_1,\dots,X_{t-1})$, the past erasure history $E^{t-1}$, and any additional side information useful for predicting the next symbol $X_t $. The predictor is treated as a black box and may vary with time~$t$, allowing for online adaptation. No assumptions are made regarding its predictive accuracy.  

For a symbol sequence $X^T=(X_1,\dots,X_T)$ of length $T$, the quality of the reconstructed sequence $\hat{X}^T=(\hat{X}_1,\dots,\hat{X}_T)$ is evaluated using a bounded per-symbol distortion  function $d: \mathcal{X} \times \mathcal{X} \to [0, D_{\max}]$ with $d(X, X) = 0$. For a given target distortion level $D \in [0, D_{\max}]$, the objective is to design the encoding and decoding sequences, $\{g_t^{\rm enc}\}^T_{t=1}$ and $\{g_t^{\rm dec}\}^T_{t=1}$, to minimize the average bit rate  
\begin{align}\label{eqn:avg_compression_rate}
    R_T = \frac{1}{T} \sum_{t=1}^T b_t,
\end{align}
while at the same time satisfying the deterministic distortion guarantee
\begin{align}
    \label{eq:finite_sample_dist_guarantee}
    \frac{1}{T} \sum_{t=1}^T d( X_t,\hat{X}_t) \le D + \frac{C}{T^{\gamma}}, 
\end{align}
for some constants $C > 0$ and $\gamma>0$. The bound \eqref{eq:finite_sample_dist_guarantee} must apply uniformly over all horizons $T$ and for all source sequences $X^T$ and erasure sequences $E^T$, thereby implying the asymptotic requirement  
\begin{align}
    \label{eq:dist_guarantee}
    \limsup_{T \to \infty} \frac{1}{T} \sum_{t=1}^T d( X_t,\hat{X}_t) \le D.
\end{align}

We note that, in practical wireless systems, the rate (\ref{eqn:avg_compression_rate}) is a proxy for the actual 
average rate, since the actual transmission rate over the wireless channel depends on the packetization and the scheduling policy.

In the following, we tackle this design problem in stages, considering first noiseless channels, i.e., $E_t=0$ for all times $t=1,\dots,T$, in Section \ref{sec:occ_error_free} and Section \ref{sec:ConfRD_noiseless}, while addressing the more general case of noisy erasure channels in Section \ref{sec:noisy}. Specifically, in Section \ref{sec:occ_error_free}, we study the special case of the outage distortion measure, while in Section \ref{sec:ConfRD_noiseless}, we introduce a general-purpose methodology applicable to any bounded distortion metric.
\section{Online Conformal Sparse Compression with Ideal Communication}

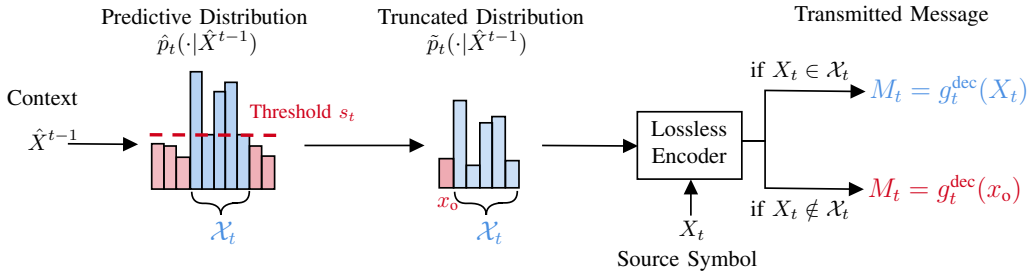
\begin{figure}[ht]
\centering
\scalebox{0.85}{\tikzset{every picture/.style={line width=0.75pt}} 

\begin{tikzpicture}[x=0.75pt,y=0.75pt,yscale=-1,xscale=1]

\draw  [fill={rgb, 255:red, 208; green, 2; blue, 27 }  ,fill opacity=0.31 ] (160.31,106.98) -- (168,106.98) -- (168,126.57) -- (160.31,126.57) -- cycle ;
\draw  [fill={rgb, 255:red, 74; green, 144; blue, 226 }  ,fill opacity=0.4 ] (118.04,56.9) -- (124.85,56.9) -- (124.85,126.57) -- (118.04,126.57) -- cycle ;
\draw  [fill={rgb, 255:red, 208; green, 2; blue, 27 }  ,fill opacity=0.31 ] (153.06,100.99) -- (160.31,100.99) -- (160.31,126.57) -- (153.06,126.57) -- cycle ;
\draw  [fill={rgb, 255:red, 74; green, 144; blue, 226 }  ,fill opacity=0.4 ] (124.85,94.27) -- (132.1,94.27) -- (132.1,126.57) -- (124.85,126.57) -- cycle ;
\draw  [fill={rgb, 255:red, 74; green, 144; blue, 226 }  ,fill opacity=0.37 ] (145.81,94.27) -- (153.06,94.27) -- (153.06,126.57) -- (145.81,126.57) -- cycle ;
\draw  [fill={rgb, 255:red, 74; green, 144; blue, 226 }  ,fill opacity=0.4 ] (138.56,63.19) -- (145.71,63.19) -- (145.71,126.57) -- (138.56,126.57) -- cycle ;
\draw  [fill={rgb, 255:red, 74; green, 144; blue, 226 }  ,fill opacity=0.4 ] (132.1,68.72) -- (138.79,68.72) -- (138.79,126.57) -- (132.1,126.57) -- cycle ;
\draw    (45.33,99.67) -- (84,99.98) ;
\draw [shift={(87,100)}, rotate = 180.46] [fill={rgb, 255:red, 0; green, 0; blue, 0 }  ][line width=0.08]  [draw opacity=0] (8.93,-4.29) -- (0,0) -- (8.93,4.29) -- cycle    ;
\draw    (415,142.32) -- (415,123.82) ;
\draw [shift={(415,120.82)}, rotate = 90] [fill={rgb, 255:red, 0; green, 0; blue, 0 }  ][line width=0.08]  [draw opacity=0] (8.93,-4.29) -- (0,0) -- (8.93,4.29) -- cycle    ;
\draw  [fill={rgb, 255:red, 208; green, 2; blue, 27 }  ,fill opacity=0.27 ] (109.67,107.67) -- (117.67,107.67) -- (117.67,126.57) -- (109.67,126.57) -- cycle ;
\draw  [fill={rgb, 255:red, 208; green, 2; blue, 27 }  ,fill opacity=0.27 ] (102.33,100.99) -- (109.59,100.99) -- (109.59,126.57) -- (102.33,126.57) -- cycle ;
\draw  [fill={rgb, 255:red, 208; green, 2; blue, 27 }  ,fill opacity=0.27 ] (95.08,99.65) -- (102.33,99.65) -- (102.33,126.57) -- (95.08,126.57) -- cycle ;
\draw [color={rgb, 255:red, 208; green, 2; blue, 27 }  ,draw opacity=1 ][line width=1.5]  [dash pattern={on 5.63pt off 4.5pt}]  (93.89,94.31) -- (171.67,94.67) ;
\draw    (185.72,100.67) -- (254,100.67) ;
\draw [shift={(257,100.67)}, rotate = 180] [fill={rgb, 255:red, 0; green, 0; blue, 0 }  ][line width=0.08]  [draw opacity=0] (8.93,-4.29) -- (0,0) -- (8.93,4.29) -- cycle    ;
\draw  [fill={rgb, 255:red, 74; green, 144; blue, 226 }  ,fill opacity=0.29 ] (274.2,74) -- (281.75,74) -- (281.75,125.82) -- (274.2,125.82) -- cycle ;
\draw  [fill={rgb, 255:red, 74; green, 144; blue, 226 }  ,fill opacity=0.29 ] (281.75,112.47) -- (289.78,112.47) -- (289.78,125.82) -- (281.75,125.82) -- cycle ;
\draw  [fill={rgb, 255:red, 74; green, 144; blue, 226 }  ,fill opacity=0.29 ] (304.97,109.69) -- (313,109.69) -- (313,125.82) -- (304.97,125.82) -- cycle ;
\draw  [fill={rgb, 255:red, 74; green, 144; blue, 226 }  ,fill opacity=0.29 ] (296.94,83.53) -- (304.86,83.53) -- (304.86,125.82) -- (296.94,125.82) -- cycle ;
\draw  [fill={rgb, 255:red, 74; green, 144; blue, 226 }  ,fill opacity=0.29 ] (289.78,87.22) -- (297.19,87.22) -- (297.19,125.82) -- (289.78,125.82) -- cycle ;
\draw  [fill={rgb, 255:red, 208; green, 2; blue, 27 }  ,fill opacity=0.33 ] (265.68,108.5) -- (274.2,108.5) -- (274.2,125.82) -- (265.68,125.82) -- cycle ;
\draw    (445.03,97.99) -- (459.03,97.99) -- (459.03,67.99) -- (515.67,68.63) ;
\draw [shift={(518.67,68.67)}, rotate = 180.65] [fill={rgb, 255:red, 0; green, 0; blue, 0 }  ][line width=0.08]  [draw opacity=0] (8.93,-4.29) -- (0,0) -- (8.93,4.29) -- cycle    ;
\draw    (445.03,97.99) -- (459.03,97.99) -- (459.03,125.99) -- (514.67,126.63) ;
\draw [shift={(517.67,126.67)}, rotate = 180.66] [fill={rgb, 255:red, 0; green, 0; blue, 0 }  ][line width=0.08]  [draw opacity=0] (8.93,-4.29) -- (0,0) -- (8.93,4.29) -- cycle    ;
\draw   (118.83,128.15) .. controls (118.9,132.82) and (121.26,135.12) .. (125.93,135.05) -- (126.02,135.05) .. controls (132.69,134.96) and (136.05,137.24) .. (136.11,141.91) .. controls (136.05,137.24) and (139.35,134.86) .. (146.01,134.77)(143.01,134.81) -- (146.1,134.77) .. controls (150.77,134.7) and (153.07,132.34) .. (153,127.67) ;
\draw   (274.83,128.15) .. controls (274.89,132.82) and (277.25,135.12) .. (281.92,135.06) -- (283.51,135.04) .. controls (290.18,134.95) and (293.54,137.24) .. (293.6,141.91) .. controls (293.54,137.24) and (296.84,134.87) .. (303.51,134.78)(300.51,134.81) -- (305.09,134.75) .. controls (309.76,134.69) and (312.06,132.33) .. (312,127.66) ;
\draw    (326.72,100.67) -- (380.66,100.98) ;
\draw [shift={(383.66,100.99)}, rotate = 180.33] [fill={rgb, 255:red, 0; green, 0; blue, 0 }  ][line width=0.08]  [draw opacity=0] (8.93,-4.29) -- (0,0) -- (8.93,4.29) -- cycle    ;

\draw (91.89,30) node [anchor=north west][inner sep=0.75pt]  [font=\small] [align=left] {\begin{minipage}[lt]{52.45pt}\setlength\topsep{0pt}
\begin{center}
$\displaystyle \hat{p}_{t}( \cdot |\hat{X}^{t-1})$
\end{center}

\end{minipage}};
\draw    (383.23,80.65) -- (445.23,80.65) -- (445.23,120.65) -- (383.23,120.65) -- cycle  ;
\draw (386.23,84.65) node [anchor=north west][inner sep=0.75pt]  [font=\small] [align=left] {\begin{minipage}[lt]{39.47pt}\setlength\topsep{0pt}
\begin{center}
Lossless\\ Encoder
\end{center}

\end{minipage}};
\draw (406.33,144) node [anchor=north west][inner sep=0.75pt]  [font=\small] [align=left] {$\displaystyle X_{t}$};
\draw (20,89) node [anchor=north west][inner sep=0.75pt]  [font=\small] [align=left] {$\displaystyle \hat{X}^{t-1}$};
\draw (227.75,18) node [anchor=north west][inner sep=0.75pt]  [font=\small] [align=left] {Truncated Distribution};
\draw (518.42,116.8) node [anchor=north west][inner sep=0.75pt]    {$\textcolor[rgb]{0.82,0.01,0.11}{M_{t} =g_t^{\text{dec}}( x_{\text{o}})}$};
\draw (448.85,49.12) node [anchor=north west][inner sep=0.75pt]  [font=\small] [align=left] {if $\displaystyle X_{t} \in \mathcal{X}_{t}$};
\draw (152,69.82) node [anchor=north west][inner sep=0.75pt]  [color={rgb, 255:red, 208; green, 2; blue, 27 }  ,opacity=1 ] [align=left] {$ $};
\draw (263,130) node [anchor=north west][inner sep=0.75pt]  [font=\small] [align=left] {$\displaystyle \textcolor[rgb]{0.82,0.01,0.11}{x_{\text{o}}}$};
\draw (128,144.4) node [anchor=north west][inner sep=0.75pt]  [color={rgb, 255:red, 74; green, 144; blue, 226 }  ,opacity=1 ]  {$\mathcal{X}_{t}$};
\draw (286,144.4) node [anchor=north west][inner sep=0.75pt]  [color={rgb, 255:red, 74; green, 144; blue, 226 }  ,opacity=1 ]  {$\mathcal{X}_{t}$};
\draw (7.75,66) node [anchor=north west][inner sep=0.75pt]  [font=\small] [align=left] {Context};
\draw (251.89,30) node [anchor=north west][inner sep=0.75pt]  [font=\small] [align=left] {\begin{minipage}[lt]{52.45pt}\setlength\topsep{0pt}
\begin{center}
$\displaystyle \tilde{p}_{t}( \cdot |\hat{X}^{t-1})$
\end{center}

\end{minipage}};
\draw (63.75,18) node [anchor=north west][inner sep=0.75pt]  [font=\small] [align=left] {Predictive Distribution};
\draw (369.75,162) node [anchor=north west][inner sep=0.75pt]  [font=\small] [align=left] {Source Symbol};
\draw (448.85,131.12) node [anchor=north west][inner sep=0.75pt]  [font=\small] [align=left] {if $\displaystyle X_{t} \notin \mathcal{X}_{t}$};
\draw (519.42,57.8) node [anchor=north west][inner sep=0.75pt]    {$\textcolor[rgb]{0.29,0.56,0.89}{M_{t} =g_t^{\text{dec}}( X_{t})}$};
\draw (475,16) node [anchor=north west][inner sep=0.75pt]  [font=\small] [align=left] {Transmitted Message};
\draw (151.75,75) node [anchor=north west][inner sep=0.75pt]  [font=\footnotesize,color={rgb, 255:red, 208; green, 2; blue, 27 }  ,opacity=1 ] [align=left] {Threshold $\displaystyle s_{t}$};

\end{tikzpicture}}
\caption{In OCSC, a source symbol $X_t$ is encoded using a truncated version of the shared predictive model's distribution $\hat{p}_t(\cdot \mid \hat{X}^{t-1})$. The predictive distribution is restricted to a high-probability set $\mathcal{X}_t$  of symbols whose probability exceeds a threshold $s_t$. The truncated distribution, augmented with a special outage symbol $x_{\rm o}$, is then encoded using a lossless encoder. If the source symbol $X_t$ lies within the high-probability set $\mathcal{X}_t$, the corresponding message is transmitted; otherwise, the outage symbol is transmitted. In the first case, the source symbol is perfectly reconstructed by inverting the lossless code, i.e. $X_t=\hat{X}_t$, whereas in the latter case the encoder and decoder agree to set the decoded symbol $\hat{X}_t$ to the maximum-likelihood symbol outside the subset $\mathcal{X}_t$. }
\label{fig:occ_diagram}
\end{figure}

\label{sec:occ_error_free}

In this section, we focus on \textit{error-free} communication channels, i.e., $E_t = 0$ for all $t \geq 1$, and on the outage distortion, or $0-1$ distortion,
\begin{align}
    \label{eq:0-1distortion}
    d(\hat{X},X) = \mathds{1}\{\hat{X}\neq X\},
\end{align}
where $\mathds{1}\{\cdot\}$ is the indicator function, i.e., $\mathds{1}\{\text{True}\}=1$ and $\mathds{1}\{\text{False}\}=0$.
Using the distortion measure \eqref{eq:0-1distortion}, we can interpret the average distortion in \eqref{eq:finite_sample_dist_guarantee} as the fraction of outages, i.e., of incorrectly decoded source symbols. In the following, we introduce \gls{ocsc}, a zero-delay compression scheme that will be shown to guarantee the requirement \eqref{eq:finite_sample_dist_guarantee} on the average outage rate.

\subsection{Online Conformal Sparse Compression}
As illustrated in Fig.~\ref{fig:occ_diagram}, \gls{ocsc} adopts a thresholding-based encoding strategy, which results in a sparse encoding distribution with a time-varying support that retains only the entries of the original predictive distribution exceeding the threshold.
Specifically, at every time $t$, only symbols that have a probability no smaller than a threshold level $s_t$ under the model's predictive distribution $\hat{p}_t(\cdot \mid \hat{X}^{t-1})$ are encoded, while the remaining symbols are associated with an outage symbol. In this way, if the source symbol $X_t$ belongs to the high-probability set $\mathcal{X}_t$, it is transmitted using a lossless code and recovered exactly at the receiver. Otherwise, an \emph{outage message} is transmitted, and both transmitter and receiver fall back to a maximum-likelihood reconstruction based on the predictor's output. By adapting the threshold level $s_t$ using an online conformal prediction rule \cite{gibbs2021adaptive}, we can ensure that \gls{ocsc} satisfies the distortion guarantee in \eqref{eq:finite_sample_dist_guarantee} for the outage distortion \eqref{eq:0-1distortion}.

In the general setting of Fig.~\ref{fig:occ_diagram}, \gls{ocsc} uses as context $Y_t$ the past reconstructed sequence, i.e., $Y_t = \hat{X}^{t-1}$, which is, by design, available at both the encoder and decoder. Specifically, based on the predictive distribution $\hat{p}_t(\cdot \mid \hat{X}^{t-1})$ and a threshold level $s_t \geq 0$, the encoder defines the \textit{high-probability set }
\begin{align}
    \label{eq:high_prob_set}
    \mathcal{X}_t = \left\{x \in \mathcal{X} : \hat{p}_t(x \mid \hat{X}^{t-1}) \geq s_t \right\}
\end{align}
to serve as support of the encoding distribution.

More precisely, for a target outage distortion level $D\in[0,1]$, \gls{ocsc} encodes the source symbol $X_t$ using an \emph{augmented truncated distribution}, which is defined over the extended alphabet $ \tilde{\mathcal{X}}_t = \mathcal{X}_t \cup \{x_{\rm o}\}$. The symbol $x_{\rm o}\notin \mathcal{X}$ is a special \emph{outage symbol} used to account for the possibility that the source symbol is not in the high-probability set, i.e., $X_t \notin \mathcal{X}_t$. Formally, the augmented truncated distribution is defined as
\begin{align}
    \label{eq:truncated}
    \tilde{p}_t(x \mid \hat{X}^{t-1}) =\begin{cases}
         (1-D) \frac{\hat{p}_t(x \mid \hat{X}^{t-1})}{\sum_{x'\in\mathcal{X}_t} \hat{p}_t(x' \mid \hat{X}^{t-1})},\quad  &x \in \mathcal{X}_t, \\
          D,\quad  &x=x_{\rm o}.
    \end{cases}
\end{align}
In words, the distribution \eqref{eq:truncated} is proportional to the predictive distribution on the high-probability set $\mathcal{X}_t$, while assigning probability $D$ to the erasure symbol $x_{\rm o}$. 

For a given source symbol $X_t\in\mathcal{X}$, the transmitter encodes the symbol 
\begin{align}
    \tilde{X}_t=\begin{cases}
         X_t,\quad  &X_t \in \mathcal{X}_t, \\
        x_{\rm o},\quad  &X_t \notin \mathcal{X}_t
    \end{cases}
\end{align} 
by using a lossless prefix-free entropy code based on the distribution \eqref{eq:truncated}. Decoding is performed by inverting the lossless prefix-free entropy code. This way, if $\tilde{X}_t \in \mathcal{X}_t$, the receiver reconstructs the source symbol exactly by decoding $\hat{X}_t=\tilde{X}_t$. Otherwise, if $\tilde{X}_t=x_{\rm o}$, the decoder outputs the maximum-likelihood symbol outside the high-probability set $\mathcal{X}_t$ under the predictor's distribution, i.e.,
\begin{align}
    \label{eq:ml_occ}
    \hat{X}_t = \argmax_{x \notin \mathcal{X}_t} \hat{p}_t(x \mid \hat{X}^{t-1}),
\end{align}
where ties are broken according to any pre-defined criterion. When $\tilde{X}_t=x_{\rm o}$, the decoded symbol \eqref{eq:ml_occ} is also evaluated at the transmitter to maintain a synchronized sequence $\hat{X}^{t}=(\hat{X}_{1},\dots,\hat{X}_{t})$. Overall, symbols in the high-probability set $\mathcal{X}_t$ are recovered without error, while those outside the set may lead to reconstruction errors due to maximum-likelihood decoding.

In the definition of the high-probability set \eqref{eq:high_prob_set}, the threshold level $s_t$ is a tunable hyperparameter that allows trading off distortion and bit rate. In fact, at every time step $t$, a larger value of the threshold $s_t$ leads to more aggressive truncation of the predictive distribution $\hat{p}_t(\cdot \mid \hat{X}^{t-1})$ and to shorter codewords, at the cost of an increase of the likelihood of an outage event. \gls{ocsc} adapts the threshold level $s_t$ using an online conformal prediction policy \cite{gibbs2021adaptive} to achieve any desired distortion level $D \in [0,1]$ in \eqref{eq:finite_sample_dist_guarantee}. Specifically, at each time $t$, \gls{ocsc} sets the threshold level $s_t$ as
\begin{align}
    \label{eq:threshold_level}
    s_t = \max\{0, \lambda_t\},
\end{align}
where the parameter $\lambda_t$ evolves according to the update rule
\begin{align}
    \label{eq:update_rule_occ_noiseless}
    \lambda_{t+1} = \lambda_{t} - \eta \left(\mathds{1}\left\{X_{t}\notin \mathcal{X}_{t}\right\} - D\right),
\end{align}
with step size $\eta > 0$ and initial value $\lambda_0 > 0$. Through the update rule \eqref{eq:update_rule_occ_noiseless}, if an outage occurs at time $t$, i.e., $X_t\notin \mathcal{X}_t$, the parameter $\lambda_{t-1}$ is decreased by a factor $\eta(1-D)$, thereby reducing the threshold level $s_t$ in \eqref{eq:threshold_level} potentially down to $0$. 
This decrease in the threshold level allows symbols with lower predictive probability to be encoded, thereby reducing the likelihood of an outage event. In contrast, if the current source symbol lies in the high-probability set $\mathcal{X}_t$, the source symbol is perfectly reconstructed, and the update \eqref{eq:update_rule_occ_noiseless} results in an increase of the threshold level by $\eta D$, which yields a more aggressive compression in future steps.

\subsection{Theoretical Guarantees}
For every target distortion level $D\in[0,1]$, \gls{ocsc} satisfies the following deterministic distortion guarantee.

\begin{theorem}
\label{th:occ_guarantee_noiseless}
Given a noiseless channel, i.e., $\{E_t=0\}^T_{t=1}$, for every sequence of source symbols $\{X_t\}_{t=1}^T$, every sequence of predictors $\{\hat{p}_t\}_{t=1}^T$, and every target distortion level $D \in [0, 1]$, the distortion of the reconstructed sequence $\{\hat{X}_t\}_{t=1}^T$ under \gls{ocsc} satisfies 
\begin{align}
    \frac{1}{T} \sum_{t=1}^T  \mathds{1}\{\hat{X}_t\neq X_t\} \leq D + \frac{ \eta(1 - D)  +\lambda_0}{\eta T}.
\end{align}
\end{theorem}
\begin{proof}
    See Appendix \ref{app:occ_guarantee_noiseless}.
\end{proof}

    Theorem \ref{th:occ_guarantee_noiseless} stipulates that \gls{ocsc} satisfies the distortion requirement \eqref{eq:finite_sample_dist_guarantee} with $\gamma=1$ and $C=(\eta(1 - D)  +\lambda_0)/\eta$. Note that the constant $C$ depends on the step size $\eta$, the initial value of the hyperparameter $\lambda_0$, and the target distortion $D$. 
\section{Online Conformal Rate-Distortion Compression with Ideal Communication}

\begin{figure}[ht]
\centering
\scalebox{0.85}{\input{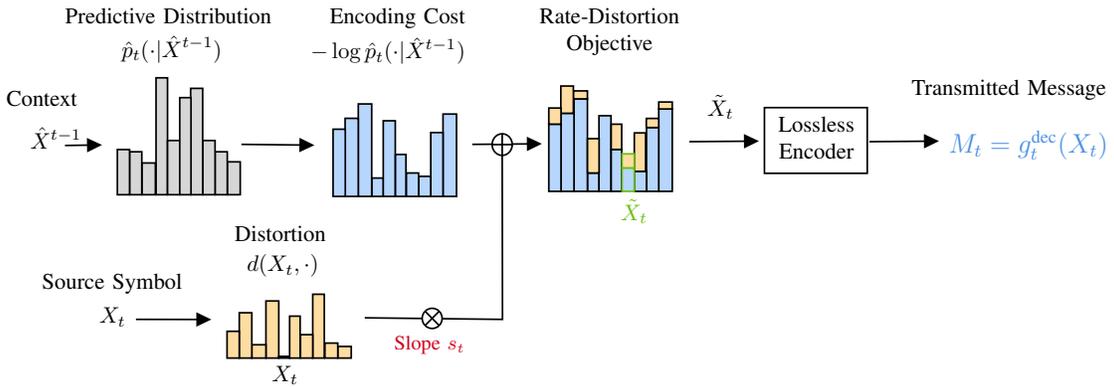}}
\label{fig:confrd_diagram}
\caption{In OCRDC, a source symbol $X_t$ is mapped to a transmitted symbol $\tilde{X}_t$ by minimizing a rate-distortion objective. The objective \eqref{eq:rd_obj} is defined as the sum of the encoding cost under the predictive distribution given the context $\hat{X}^{t-1}$, and of the distortion cost, which is computed using the source symbol $X_t$. The distortion term is scaled by a slope parameter $s_t$. The symbol $\tilde{X}_t$ that minimizes the rate-distortion objective is transmitted using a lossless code. Decoding is performed by inverting the lossless code.}
\end{figure}
\label{sec:ConfRD_noiseless}

As detailed in the previous section, \gls{ocsc} exploits the unique ``all-or-nothing'' property of the outage distortion to design an encoding scheme that controls the support of the encoding distribution as a function of the target distortion. More general distortion measures do not allow for this special design. For this reason, in this section, we introduce \textit{\gls{ocrdc}}, a general-purpose lossy compression scheme that satisfies deterministic distortion guarantees for every bounded distortion measure.

\subsection{Online Conformal Rate-Distortion Compression}

\Gls{ocrdc} encodes the source symbols by minimizing a weighted rate-distortion objective, which is defined using the predictive distribution $\hat{p}_t(\cdot \mid Y_{t})$, with the context $Y_{t}$ given by the past reconstructed symbols, i.e. $Y_{t} = \hat{X}^{t-1}$. Unlike \gls{ocsc}, which controls a threshold parameter $s_t$, \gls{ocrdc} maintains a \textit{slope parameter} $s_t$, which is used to instantiate the weighted rate-distortion objective.

Specifically, at each round $t$, given a slope value $s_t \geq 0$, the encoder maps the source symbol $X_t$ to a transmitted symbol $\tilde{X}_t$ by minimizing the rate-distortion objective \cite{berger2003rate}
\begin{align}
    \label{eq:rd_obj}
    \tilde{X}_t = \argmin_{\tilde{X} \in \mathcal{X}} \left\{ -\log \hat{p}_t(\tilde{X}\mid\hat{X}^{t-1}) + s_t \, d(X_t, \tilde{X}) \right\}.
\end{align}
The objective in \eqref{eq:rd_obj}, routinely used in the rate-distortion literature \cite{berger2003rate,jalali2011lossy,dai2022nonlinear}, consists of two terms: the first term represents the code length of symbol $\tilde{X}$ under the predictive distribution $\hat{p}_t(\tilde{X}\mid\hat{X}^{t-1})$; in contrast, the second term, weighted by the slope value $s_t$, quantifies the distortion incurred when replacing the source symbol $X_t$ with $\tilde{X}$. Unlike conventional static schemes \cite{yang1997fixed,jalali2008rate}, which keep the slope $s_t$ unchanged over time, \gls{ocrdc} updates the slope via an online conformal prediction rule to be discussed next.

The selected symbol $\tilde{X}_t$ in \eqref{eq:rd_obj} is encoded by using a lossless prefix-free entropy code matched to the shared distribution $\hat{p}_t(\tilde{X}\mid\hat{X}^{t-1})$. We note that, unlike the scheme in the previous section, in this scheme the communication parties do not need to threshold the predictive distribution. The resulting message $M_t$ is transmitted to the decoder, which receives the message $\tilde{M}_t=M_t$ error-free. The lossless prefix-free entropy code is then inverted, and the receiver reconstructs the transmitted symbol as $\hat{X}_t=\tilde{X}_t$.

In \gls{ocrdc}, the slope value $s_t$ in \eqref{eq:rd_obj} is used to trade off distortion and bit rate: larger values of $s_t$ prioritize reduced distortion at the cost of longer codewords, while smaller values allow for more aggressive compression with potentially higher distortion. With the aim of satisfying the requirement \eqref{eq:finite_sample_dist_guarantee}, \gls{ocrdc} defines the slope value $s_t$ as in \eqref{eq:threshold_level}, where the parameter $\lambda_t$ is updated over time as 
\begin{align}
    \label{eq:update_rule_noiseless}
    \lambda_{t+1} = \lambda_{t} + \eta\big(d(X_{t}, \tilde{X}_{t}) - D\big),
\end{align}
for a step size $\eta > 0$ and an initial parameter value $\lambda_0>0$. Through the rule \eqref{eq:update_rule_noiseless}, the parameter $\lambda_t$ increases when the instantaneous distortion $d(X_{t-1},\tilde{X}_{t-1})$ between the source symbol $X_{t-1}$ and the transmitted symbol $\tilde{X}_{t-1}$ exceeds the target distortion level $D$, thereby placing greater emphasis on distortion control in the next round. Conversely, when the distortion $d(X_{t-1},\tilde{X}_{t-1})$ falls below $D$, the update decreases the parameter $\lambda_t$, thereby placing more emphasis on bit rate reduction.  

\subsection{Theoretical Guarantees}

We now show that \gls{ocrdc} also meets the requirement \eqref{eq:finite_sample_dist_guarantee} under the following technical assumption. 
To introduce it, note that in order to ensure that, for sufficiently large values of the slope value $s_t$, the source symbol can be recovered with arbitrarily low distortion, the encoding cost of the source symbol $X_t$ in \eqref{eq:rd_obj}, given by $-\log \hat{p}_t(\tilde{X}\mid\hat{X}^{t-1})$, must remain finite. 
Since our results hold in an adversarial setting where the source sequence may be chosen arbitrarily, possibly depending on the predictive distribution, we require that the encoding cost $-\log \hat{p}_t(x\mid\hat{X}^{t-1})$ remain bounded for every symbol $x \in \mathcal{X}$. In practice, this condition can be enforced by smoothing the predictive distribution, for example, by mixing $\hat{p}_t(\cdot\mid\hat{X}^{t-1})$ with a uniform distribution over the source domain $\mathcal{X}$.

\begin{assumption}
\label{ass:max_cost}
There exists a constant $L < \infty$ such that for all $t \geq 1$, the predictive model distribution $\hat{p}_t(\cdot\mid\hat{X}^{t-1})$ satisfies the inequality
\begin{align}
    \label{eq:max_len}
    \max_{x \in \mathcal{X}} \{-\log \hat{p}_t(x\mid\hat{X}^{t-1})\} < L.
\end{align}
\end{assumption}
Note that the left-hand side of \eqref{eq:max_len} is the Rényi entropy
of order $\infty$ for the predictive distribution $\hat{p}_t(\cdot\mid\hat{X}^{t-1})$ \cite{Simeone_2025}. Therefore, the assumption \eqref{eq:max_len} imposes a lower bound on the uncertainty of the predictive distribution $\hat{p}_t(\cdot\mid\hat{X}^{t-1})$.
Under Assumption \ref{ass:max_cost}, it is possible to show that the value of the parameter $\lambda_t$ remains bounded over time.
\begin{lemma}
\label{lemma:bounded_lambda}
Under Assumption \ref{ass:max_cost}, for a target distortion $D>0$, the parameter  $\lambda_t$ generated by the update rule \eqref{eq:update_rule_noiseless} satisfies the inequality
\begin{align}
    \lambda_t \leq \frac{L}{D} + \eta(D_{\rm max} - D),
\end{align}
for all $t\geq1 $.
\end{lemma}
\begin{proof}
    See Appendix \ref{proof:bounded_lambda}.
\end{proof}
Following standard arguments from online conformal prediction \cite{gibbs2021adaptive,angelopoulos2024online}, we establish next that \gls{ocrdc} satisfies the long-term distortion guarantee \eqref{eq:finite_sample_dist_guarantee}.

\begin{theorem}
\label{th:distortion_noiseless}
Given a noiseless channel, i.e., $\{E_t=0\}^T_{t=1}$, for any sequence of source symbols $\{X_t\}_{t=1}^T$, any sequence of predictors $\{\hat{p}_t\}_{t=1}^T$ satisfying Assumption \ref{ass:max_cost}, and any target distortion level $D \in (0, D_{\max}]$, the distortion of the reconstructed sequence $\{\hat{X}_t\}_{t=1}^T$ under \gls{ocrdc} satisfies the requirement \eqref{eq:finite_sample_dist_guarantee} as
\begin{align}
    \label{eq:ocrdc_guarantee}
    \frac{1}{T} \sum_{t=1}^T d( X_t,\hat{X}_t)
    \leq D + \frac{L/D + \eta(D_{\max} - D) - \lambda_0}{\eta T}.
\end{align}
\end{theorem}
\begin{proof}
    See Appendix \ref{proof:distortion_noiseless}.
\end{proof}
Theorem \ref{th:distortion_noiseless} states that, under the condition of Assumption \ref{ass:max_cost}, \gls{ocrdc} meets the distortion requirement  \eqref{eq:finite_sample_dist_guarantee} with $\gamma=1$ and $C=(L/D + \eta(D_{\max} - D) - \lambda_0)/\eta$. Note that the constant $C$ depends on the algorithm's hyperparameters, such as the step size $\eta$, the initial value $\lambda_0$, the bound on the predictive distribution $L$, the maximum distortion $D_{\rm max}$ and the target distortion $D$. While Theorem \ref{th:distortion_noiseless} does not cover the case of lossless encoding, i.e., $D=0$, the guarantee in \eqref{eq:ocrdc_guarantee} can be satisfied for $D=0$ by simply setting $s_t = \infty$ in \eqref{eq:rd_obj}, ensuring that $\tilde{X}_t = X_t$ for all $t \geq 1$.

\section{Channel-Adaptive Online Conformal Compression for Erasure Channels}
\label{sec:noisy}
In this section, we extend \gls{ocsc} and \gls{ocrdc} to communication channels subject to errors in the form of erasures. In particular, we develop a strategy to adapt their operation to account for the distortion induced by channel erasures, establishing sufficient conditions on the channel under which the distortion requirements \eqref{eq:finite_sample_dist_guarantee} and \eqref{eq:dist_guarantee} can be met. We also derive explicit distortion guarantees for deterministic, as well as for stochastic erasure channels with and without memory. Unless stated otherwise, the techniques and results presented below apply to both \gls{ocsc} and \gls{ocrdc}. In this section, the hyperparameter $s_t$ should be interpreted as a threshold in \gls{ocsc} and as a slope in \gls{ocrdc}. The two schemes are referred to collectively as \textit{online conformal compression} (OCC).
In the absence of erasures, the encoder operates as illustrated in Sections \ref{sec:occ_error_free} and \ref{sec:ConfRD_noiseless} for \gls{ocsc} and \gls{ocrdc}, respectively. In contrast, in the case of an erasure, we assume that the decoder maps the message $\tilde{M}_t = \circledast$ in \eqref{eq:channel_output}  to the maximum-likelihood decoding estimate with respect to the predictive distribution
\begin{align}
    \label{eq:maximum_likelihood}
    \hat{X}_t = \argmax_{x \in \mathcal{X}} \hat{p}_t(x \mid \hat{X}^{t-1}).
\end{align}
With this choice, since the erasure indicator $E_t$ is fed back to the encoder at time $t+1$, the encoder can reconstruct the decoded symbol $\hat{X}_t$ in \eqref{eq:maximum_likelihood} even in the event of an erasure. Therefore, the encoder can operate as described in the previous sections for both \gls{ocsc} and \gls{ocrdc}.

\subsection{Impact of Channel Erasures}
Channel erasures introduce an additional source of distortion to that introduced by lossy encoding at the transmitter. We define the \textit{channel-induced distortion} at time $t$, denoted as $\delta^{\rm ch}_t$, as the difference between the distortion incurred when decoding based on the actual received message $\tilde{M}_t$ in \eqref{eq:channel_output}, i.e., $d(X_t, g^{\rm dec}_t(\tilde{M}_t))$, and the distortion $d(X_t, g^{\rm dec}_t(M_t))$ that would have been incurred if the transmitted message $M_t$ was received without error. This is given by the difference
\begin{align}
    \label{eq:distortion_gap}
    \delta^{\rm ch}_t = d(X_t, g^{\rm dec}_t(\tilde{M}_t)) - d(X_t, g^{\rm dec}_t(M_t)).
\end{align}
For \gls{ocsc}, the channel-induced distortion $\delta^{\rm ch}_t$ is binary, and it equals $\delta^{\rm ch}_t = 1$ only when an erasure occurs, and the source $X_t$ differs from the maximum-likelihood symbol \eqref{eq:maximum_likelihood}, and it would be correctly reconstructed in the absence of erasure, i.e., $g^{\rm dec}_t(M_t)=X_t$. In contrast, for \gls{ocrdc}, the channel-induced distortion is given by the difference between the distortion of the reconstructed symbol $\hat{X}_t$ and that of the transmitted symbol $\tilde{X}_t$. Overall, we have
\begin{align}
    \label{eq:distortion_gap_true}
    \delta^{\rm ch}_t =
\begin{cases}
E_t \cdot \mathds{1}\!\left\{(X_t=g^{\rm dec}_t(M_t)) \cap \left({X}_t \neq \argmax_{x \in \mathcal{X}} \hat{p}_t(x \mid \hat{X}^{t-1})\right)\right\}, & \text{for OCSC,}  \\
d(X_t, \hat{X}_t) - d(X_t, \tilde{X}_t), & \text{for \gls{ocrdc}.}
\end{cases}
\end{align}
The channel-induced distortion in \eqref{eq:distortion_gap_true} is non-negative. In fact, when $E_t = 0$, there is no error and we have $\delta^{\rm ch}_t = 0$, whereas, if $E_t = 1$ and an erasure occurs, then we have $\delta^{\rm ch}_t \geq 0$. This inequality is straightforward for \gls{ocsc}, while, for \gls{ocrdc}, it follows because the erasure-free reconstruction $\tilde{X}_t$ is the solution to the optimization problem \eqref{eq:rd_obj}, while the actual reconstructed symbol $\hat{X}_t$ is the maximum-likelihood estimate. 

The channel-induced distortion $\delta^{\rm ch}_t$ can be evaluated at the transmitter at time $t+1$, since it has access to both $\tilde{X}_t$ and $\hat{X}_t$. The receiver, instead, cannot evaluate it because the transmitted symbol $\tilde{X}_t$ is unknown in the case of an erasure.  This information asymmetry implies that the history of channel distortion $\delta^{\rm ch}_1, \dots, \delta^{\rm ch}_t$ cannot be directly used to design the threshold level $s_t$ in OCSC. Indeed, in \gls{ocsc}, both the encoder and decoder must be fed with the same threshold $s_t$. In contrast, for \gls{ocrdc}, this is not problematic, since only the encoder operation depends on the slope level $s_t$. For this reason, we  define the following upper bound on the channel-induced distortion that can be used in the design of the hyperparameter $s_t$ for both schemes,
\begin{align}
    \label{eq:distortion_gap_surrogate}
    \delta^{\rm ch}_t\leq\tilde\delta^{\rm ch}_t =
\begin{cases}
E_t , & \text{for OCSC,}  \\
d(X_t, \hat{X}_t) - d(X_t, \tilde{X}_t), & \text{for \gls{ocrdc}.}
\end{cases}
\end{align}
Note that for \gls{ocrdc}, the upper bound is tight, i.e., $\delta^{\rm ch}_t=\tilde\delta^{\rm ch}_t$. 

We denote the corresponding upper bound on the cumulative channel-induced distortion up to time $t$ as
\begin{align}
    \Delta^{\rm ch}_t = \sum_{i=1}^t \tilde\delta_i^{\rm ch}.
\end{align}
Thus, the quantity $\Delta^{\rm ch}_t$ provides an upper bound on the additional distortion in the reconstructed sequence $\hat{X}^t$ that occurs solely due to channel erasures. 

\subsection{Channel-Adaptive Online Conformal Rate-Distortion Compression}
To compensate for the increased distortion resulting from channel errors, we consider a doubly-adaptive variant of the update rules \eqref{eq:update_rule_occ_noiseless} and \eqref{eq:update_rule_noiseless}, which we term \textit{Channel-Adaptive OCSC}  (CA-OCSC) and \textit{Channel-Adaptive OCRDC } (CA-OCRDC). 
At time $t$, the hyperparameter $s_t$ is defined as \eqref{eq:threshold_level}, where the parameter $\lambda_t$ is updated following the doubly-adaptive update rule

\begin{align}
    \label{eq:update_rule_noisy}
    \lambda_{t+1} &=\begin{cases}
\lambda_{t} - \eta \left((1-E_t)\mathds{1}\{X_t \notin \mathcal{X}_t\} - D + \delta^{\rm tgt}_t\right), & \text{for CA-OCSC,}  \\
\lambda_t + \eta (d(X_t, \tilde{X}_t) - D + \delta^{\rm tgt}_t ), & \text{for CA-OCRDC,}
\end{cases} 
\end{align}
with
\begin{align}
    \delta^{\rm tgt}_{t+1} &= \min\big\{ D - \epsilon, \; \Delta^{\rm ch}_t - \Delta^{\rm tgt}_t \big\},
    \label{eq:adjustment_policy}
\end{align}
where $\Delta^{\rm tgt}_t$ is the cumulative target distortion adjustment
\begin{align}
    \Delta^{\rm tgt}_t = \sum_{i=1}^t \delta^{\rm tgt}_i.
\end{align}
In \eqref{eq:adjustment_policy}, the parameter $\epsilon>0$ ensures that the adjusted distortion requirement $D-\delta^{\rm tgt}_t $ remains positive and strictly larger than $\epsilon$. Imposing the strict inequality $\epsilon>0$ is necessary to obtain a non-vacuous bound on the value of $\lambda_t$ in Lemma \ref{lemma:bounded_lambda_timevarying}.

The key difference between the update rule \eqref{eq:update_rule_noisy} for noisy channels and the corresponding update rules \eqref{eq:update_rule_occ_noiseless} and \eqref{eq:update_rule_noiseless} for error-free channels is the adjustment of the target distortion level by the factor $\delta^{\rm tgt}_t$, which is designed to compensate for channel-induced distortion. Specifically, the adjustment $\delta^{\rm tgt}_{t+1}$ in \eqref{eq:update_rule_noisy} is the residual \eqref{eq:adjustment_policy} between the cumulative channel distortion $\Delta^{\rm ch}_t$ and the cumulative target distortion adjustment $\Delta^{\rm tgt}_t$, which is capped to $D - \epsilon$. 

The design of the target distortion adjustment \eqref{eq:adjustment_policy} is inspired by a \textit{queueing-theoretic} analogy, where the upper bound on the channel distortion $\tilde\delta^{\rm ch}_t$ acts as an arrival process and the adjustment $\delta^{\rm tgt}_t$ acts as a service process \cite{neely2010stochastic}. The update rule \eqref{eq:adjustment_policy} can be viewed as a service policy aimed at draining the residual distortion queue
\begin{align}
    \label{eq:distortion_queue}
    Q_t = \Delta^{\rm ch}_t - \Delta^{\rm tgt}_t,
\end{align}
subject to a maximum service rate of $D - \epsilon$.

\subsection{Theoretical Guarantees}

We now show that CA-OCC, i.e., both CA-OCSC and CA-OCRDC satisfy the requirement \eqref{eq:finite_sample_dist_guarantee} under the following regularity assumption on the predictive distribution, which mirrors Assumption~\ref{ass:max_cost}.
\begin{assumption}
\label{ass:max_cost_2}
For CA-OCSC, the predictive model distribution $\hat{p}_t(\cdot \mid \hat{X}^{t-1})$ is arbitrary, whereas for CA-OCRDC, there exists a constant $L < \infty$ such that for all $t \geq 1$, the predictive model distribution $\hat{p}_t(\cdot \mid \hat{X}^{t-1})$ satisfies
\begin{align}
    \label{eq:max_len_2}
    \max_{x \in \mathcal{X}} \{-\log \hat{p}_t(x \mid \hat{X}^{t-1})\} < L.
\end{align}
\end{assumption}
Under this Assumption, the next lemma establishes that the sequence $\{\lambda_t\}_{t\geq 1}$ remains bounded for any $\epsilon > 0$.
\begin{lemma}
\label{lemma:bounded_lambda_timevarying}
For any sequence of source symbols $\{X_t\}_{t=1}^T$ and erasures $\{E_t\}_{t=1}^T$, the parameter sequence $\{\lambda_t\}_{t \geq 1}$ generated by the update \eqref{eq:update_rule_noisy} satisfies, for all $t \geq 1$,
\begin{align}
    \lambda_t \geq -\eta (1 - \epsilon), &\quad \text{for CA-OCSC}, \\
    \lambda_t \leq \tfrac{L}{\epsilon} + \eta (D_{\max} - \epsilon), &\quad \text{for CA-OCRDC}.
\end{align}
\end{lemma}
\begin{proof}
See Appendix \ref{app:bounded_lambda_timevarying}.
\end{proof}

The following proposition highlights the relationship between the distortion of the received sequence $\hat{X}^T$ and the residual distortion queue size $Q_T$ in \eqref{eq:distortion_queue}.

\begin{proposition}
    \label{prop:distortion_queue}
    For any sequence of source symbols $\{X_t\}_{t=1}^T$, any sequence of predictors $\{\hat{p}_t\}_{t=1}^T$ satisfying Assumption \ref{ass:max_cost_2}, and any target distortion level $D \in (0, D_{\max}]$, the average distortion of the reconstructed sequence $\{\hat{X}_t\}_{t=1}^T$ obtained under the parameter sequence $\{s_t\}_{t \geq 1}$ defined by \eqref{eq:threshold_level} and \eqref{eq:update_rule_noisy}-\eqref{eq:adjustment_policy} satisfies
\begin{align}
    \label{eq:intermediate_bound}
    \frac{1}{T} \sum_{t=1}^T d(X_t, \hat{X}_t) 
    \leq D + \frac{K}{\eta T} + \frac{Q_T}{T},
\end{align}
where
\begin{align}
    \label{eq:definition_K}
    K = \begin{cases}
        \eta (1 - \epsilon) + \lambda_0, & \text{for CA-OCSC}, \\
        L/\epsilon + \eta (D_{\max} - \epsilon) - \lambda_0, & \text{for CA-OCRDC}.
    \end{cases}
\end{align}
\end{proposition}
\begin{proof}
    See Appendix \ref{app:distortion_queue}.
\end{proof}
Proposition \ref{prop:distortion_queue} states that for any sequence of source symbols $\{X_t\}_{t=1}^T$, the distortion of the reconstructed sequence does not exceed the target distortion $D$, up to two additive terms: one that decays as $1/T$ and depends on the algorithm's hyperparameters, and another term that is proportional to the residual queue $Q_T$. Under ideal communication, we have $Q_T = 0$ and Proposition \ref{prop:distortion_queue} reduces to Theorem \ref{th:occ_guarantee_noiseless} and Theorem \ref{th:distortion_noiseless}. In the following sections, we establish sufficient conditions on the erasure channel under which $Q_T$ grows sublinearly in $T$, thereby ensuring the distortion guarantee \eqref{eq:finite_sample_dist_guarantee}.

\subsection{Deterministic Erasure Channels}

We first consider deterministic erasure channels, i.e., channels for which the erasure sequence $\{E_t\}_{t \geq 1}$ is deterministic. In a wireless system, this models the situation in which the system's scheduling strategy deterministically drops some of the symbols. To characterize the conditions on the erasure sequence $\{E_t\}_{t \geq 1}$ under which the residual queue $Q_T$ is sublinear and the distortion guarantee \eqref{eq:finite_sample_dist_guarantee} holds, we define the \emph{minimum envelope process} of a sequence \cite{chang1994stability}.

\begin{definition}[Minimum Envelope Process \cite{chang1994stability}]
Given a non-negative sequence $z^{\infty}=\{z_t\}_{t \geq 1}$ and integer $\tau\geq 1$, its minimum envelope process is defined as
\begin{align}
    \label{eq:distortion_mep}
    \Psi(\tau,z^{\infty}) = \sup_{k \geq 1} \sum_{t=k}^{k+\tau} z_t.
\end{align}
\end{definition}
The minimum envelope process $ \Psi(\tau,z^{\infty})$ represents the maximum cumulative sum of $z^{\infty}$ over any interval of length $\tau$. 
For the erasure sequence $E^{\infty} = \{E_t\}_{t \geq 1}$, the minimum envelope process $ \Psi(\tau,E^{\infty})$ measures the maximum number of erasures within any window of $\tau$ consecutive communication slots. In Assumption \ref{ass:bounded_channel_distortion} below, we require that the minimum envelope process of the erasure sequence be bounded by an affine function with slope strictly smaller than $(D - \epsilon)/D_{\rm max}$ for some $\epsilon>0$. This condition is sufficient to establish that the quantity $Q_T$ grows sublinearly in $T$, which is a key requirement for proving Theorem  \ref{th:general_dist_guarantee_noisy} below.
\begin{assumption}
\label{ass:bounded_channel_distortion}
There exist constants $A <(D - \epsilon)/D_{\rm max}$ and a sublinear function $\psi(\tau)$ (i.e. a function satisfying $\lim_{\tau \to \infty}\psi(\tau)/\tau= 0$) such that the minimum envelope process of the erasure sequence $ E^{\infty}=\{E_t\}_{t \geq 1}$ satisfies
\begin{align}
    \label{eq:distortion_cond}
    \Psi(\tau, E^{\infty}) < A \tau + \psi(\tau)
\end{align}
for all $\tau \geq 1$.
\end{assumption}

The following theorem shows that, in the presence of erasures, as long as the minimum envelope of the erasure sequence satisfies Assumption \ref{ass:bounded_channel_distortion}, the doubly-adaptive update rule \eqref{eq:update_rule_noisy}-\eqref{eq:adjustment_policy} guarantees the distortion requirement \eqref{eq:finite_sample_dist_guarantee} for any source symbol sequence and predictors.

\begin{theorem}
\label{th:general_dist_guarantee_noisy}
Given a deterministic erasure channel with erasure sequence $\{E_t\}_{t \geq 1}$ satisfying Assumption \ref{ass:bounded_channel_distortion}, any sequence of source symbols $\{X_t\}_{t=1}^T$, any sequence of predictors $\{\hat{p}_t\}_{t=1}^T$ satisfying Assumption \ref{ass:max_cost_2}, and a target distortion level $D \in (0, D_{\max}]$, the distortion of the reconstructed sequence $\{\hat{X}_t\}_{t=1}^T$ obtained under the parameter sequence $\{s_t\}_{t \geq 1}$ defined by \eqref{eq:threshold_level} and \eqref{eq:update_rule_noisy}-\eqref{eq:adjustment_policy} satisfies
\begin{align}
    \frac{1}{T} \sum_{t=1}^T d(X_t, \hat{X}_t) 
    \leq D + \frac{K}{T \eta} 
    + \frac{\tau_{\rm max}(D_{\rm max} - D + \epsilon)+D-\epsilon}{T},
\end{align}
where $\tau_{\rm max} < \infty$ is defined as
\begin{align}
    \label{eq:tau_max}
    \tau_{\rm max} = \min\left\{\tau \in \mathbb{N} : \frac{\psi(\tau)}{\tau} \leq \frac{D - \epsilon}{D_{\rm max}} - A \right\},
\end{align}
and $K$ as in \eqref{eq:definition_K}.
\end{theorem}

\begin{proof}
See Appendix \ref{app:general_dist_guarantee_noisy}.
\end{proof}

\subsection{Time-varying Erasure Channels}

We now consider the case of memoryless time-varying erasure channels, where the channel is described by a sequence of erasure probabilities $\{e_t\}_{t \geq 1}$ and, at each time step $t$, the erasure variable is stochastic and given by $E_t \sim \mathrm{Bern}(e_t)$. This is a packet-level model for a wireless channel whose parameters are changing dynamically. Given the randomness of the erasure process, the distortion guarantees provided in this section hold with high probability with respect to the realization of the erasure sequence. Next, we state an assumption under which a probabilistic recovery guarantee of the same flavor as \eqref{eq:finite_sample_dist_guarantee} and \eqref{eq:dist_guarantee} can be established.
\begin{assumption}
\label{ass:bounded_erasure_mep}
There exist constants $A < (D - \epsilon)/D_{\rm max}$, and a sublinear function $\psi(\tau)$ (i.e. $\lim_{\tau \to \infty}\psi(\tau)/\tau= 0$) such that the minimum envelope process of the erasure probability sequence $e^{\infty}=\{e_t\}_{t \geq 1}$ satisfies
\begin{align}
    \label{eq:distortion_cond_2}
    \Psi(\tau, e^{\infty}) < A \tau + \psi(\tau)
\end{align}
for all $\tau \geq 1$.
\end{assumption} 

Assumption \ref{ass:bounded_erasure_mep} states that the maximum cumulative sum of the erasure probability sequence $e^{\infty}=\{e_t\}_{t \geq 1}$ over any interval of length $\tau$ grows at most linearly with slope strictly smaller than $(D - \epsilon)/D_{\max}$. Under this condition, the distortion of the reconstructed sequence satisfies the finite-sample distortion condition \eqref{eq:finite_sample_dist_guarantee} with high probability and the asymptotic distortion condition \eqref{eq:dist_guarantee} almost surely. In the following, we use $\mathcal{O}(\cdot)$ to denote big-O notation.
\begin{theorem}
\label{th:dist_guarantee_erasure_ch}

Given a time-varying erasure channel with erasure probabilities $\{e_t\}_{t \geq 1}$ satisfying Assumption \ref{ass:bounded_erasure_mep}, any sequence of source symbols $\{X_t\}_{t=1}^T$, any sequence of predictors $\{\hat{p}_t\}_{t=1}^T$ satisfying Assumption \ref{ass:max_cost_2}, and a target distortion level $D \in (0, D_{\max}]$, for any $\delta_T>0$, with probability $1-\delta_T$, the distortion of the reconstructed sequence $\{\hat{X}_t\}_{t=1}^T$ obtained under the parameter sequence $\{s_t\}_{t \geq 1}$ defined by \eqref{eq:threshold_level} and \eqref{eq:update_rule_noisy}-\eqref{eq:adjustment_policy} satisfies
\begin{align}
    \frac{1}{T} \sum_{t=1}^{T} d(X_t, \hat{X}_t) \leq D + \frac{K}{T\eta} + \mathcal{O}\left(\frac{D_{\rm max}\psi(T)}{T}+\frac{D_{\rm max}}{T}\log\left(\frac{1}{\delta_T}\right)\right),
\end{align}
where $K$ is defined as in \eqref{eq:definition_K}.

Moreover, setting $\delta_T=1/T^2$, it holds almost surely
\begin{align}
    \limsup_{T \to \infty} \frac{1}{T} \sum_{t=1}^{T} d(X_t, \hat{X}_t) \le D.
\end{align}
\end{theorem}
\begin{proof}
See Appendix \ref{app:dist_guarantee_erasure_ch}.
\end{proof}

\subsection{Gilbert-Elliott Erasure Channels}
In this section, we extend our analysis to stochastic channels with memory. Specifically, we consider the Gilbert-Elliott erasure channel, a channel model with memory commonly used to model bursty erasures \cite{gilbert1960capacity,elliott1963estimates}. In a wireless setup, this models a channel with intermittent interference, which can occur in uplink grant-free transmission. This channel is modeled by a two-state Markov chain $\{Z_t\}_{t \geq 1}$ with state space $\{\mathrm{B}, \mathrm{G}\}$, representing the \textit{Bad} and \textit{Good} states, respectively. The transition probability matrix is given by
\begin{align}
\label{eq:gb_matrix}
P = \begin{pmatrix}
p_{\mathrm{BB}} & p_{\mathrm{BG}} \\
p_{\mathrm{GB}} & p_{\mathrm{GG}}
\end{pmatrix} 
= 
\begin{pmatrix}
1 - a & a \\
b & 1 - b
\end{pmatrix},
\end{align}
and the state of the Markov chain determines the erasure probability at time $t$ as
\begin{align}
\label{eq:gb_erasure_prob}
e_t = \begin{cases}
e_{\mathrm{B}}, & \text{if } Z_t = \mathrm{B}, \\
e_{\mathrm{G}}, & \text{if } Z_t = \mathrm{G}.
\end{cases}
\end{align}
The steady-state distribution of the Markov chain is given by 
\begin{align}
\pi(\mathrm{B}) = \frac{b}{a + b}, \quad \text{and} \quad \pi(\mathrm{G}) = \frac{a}{a + b},
\end{align}
and the steady-state erasure probability is
\begin{align}
\label{eq:gb_steady_state}
\bar{e} = \pi(\mathrm{B}) e_{\mathrm{B}} + \pi(\mathrm{G}) e_{\mathrm{G}}.
\end{align}

In the Gilbert-Elliott channel, the erasure probabilities are random quantities determined by the evolution of the Markov Chain $\{Z_t\}_{t \geq 1}$. Leveraging standard concentration inequalities for reversible Markov chains \cite{paulin2015concentration}, we can establish a high-probability bound on the deviation from the steady-state erasure probability $\bar{e}$ of the average erasure probabilities over any window of size $\tau$ starting at $t \in \{1, \dots, T\}$, in terms of the spectral gap $\gamma = 1 - |1 - a - b|$ \cite{paulin2015concentration}.

\begin{lemma}
\label{eq:lemma_GB}
For a Gilbert-Elliott erasure channel with spectral gap $\gamma < 1$, for any $\delta \in (0,1]$, with probability at least $1 - \delta$, for all $\tau > 1$ and $k \in \{1, \dots, T\}$, it holds that
\begin{align}
\sum_{t=k}^{k+\tau} e_t \leq \bar{e} \, \tau +
    \sqrt{ \tau\frac{1+\gamma}{1-\gamma} 
    \log\left( \frac{\sqrt{T}\pi^2\tau^2}{\min\{\pi(\mathrm{G}),\pi(\mathrm{B})\}  6\delta}\right)}.
\end{align}
\end{lemma}

\begin{proof}
See Appendix \ref{app:lemma_GB}.
\end{proof}

Lemma \ref{eq:lemma_GB} characterizes, in a probabilistic sense, the minimum envelope process of the erasure probabilities $\{e_t\}_{t\geq 1}$, showing that it can be bounded with high probability by an affine function with slope equal to the steady-state erasure probability $\bar{e}$ plus a sublinear term in $T$ and $\tau$. In a manner similar to the time-varying erasure channel case, if the steady-state erasure probability $\bar{e}$ is smaller than $(D-\epsilon)/D_{\rm max}$, the distortion of the reconstructed sequence satisfies the finite-sample distortion condition \eqref{eq:finite_sample_dist_guarantee} with high probability and the asymptotic distortion condition \eqref{eq:dist_guarantee} almost surely.
\begin{theorem}
\label{th:theorem_GB}

Given a Gilbert-Elliott erasure channel with spectral gap $\gamma<1$ and steady-state erasure probability $\bar{e} < (D-\epsilon)/D_{\rm max}$, any sequence of source symbols $\{X_t\}_{t=1}^T$, any sequence of predictors $\{\hat{p}_t\}_{t=1}^T$ satisfying Assumption \ref{ass:max_cost_2}, and a target distortion level $D \in (0, D_{\max}]$, for any $\delta_T>0$, with probability $1-\delta_T$, the distortion of the reconstructed sequence $\{\hat{X}_t\}_{t=1}^T$ obtained under the parameter sequence $\{s_t\}_{t \geq 1}$ defined  by \eqref{eq:threshold_level} and \eqref{eq:update_rule_noisy}-\eqref{eq:adjustment_policy} satisfies
\begin{align}
    \frac{1}{T} \sum_{t=1}^{T} d(X_t, \hat{X}_t) \leq &D + \frac{K}{T\eta} +\mathcal{O}\hspace{-0.2em}\left(\hspace{-0.25em}\frac{D_{\rm max}}{\sqrt{T}}\hspace{-0.1em}\sqrt{\frac{1+\gamma}{1-\gamma} 
    \log\left( \frac{4\sqrt{T}\pi^2T^2}{\min\{\pi(\mathrm{G}),\pi(\mathrm{B})\}  \delta_T} \right) }\hspace{-0.1em}+\frac{D_{\rm max}}{T}\hspace{-0.1em}\log\left(\frac{1}{\delta_T}\right)\hspace{-0.25em}\hspace{-0.1em}\right)
\end{align}
where $K$ is defined as in \eqref{eq:definition_K}.

Moreover, setting $\delta_T=1/T^2$, it holds almost surely
\begin{align}
	\limsup_{T \to \infty} \frac{1}{T} \sum_{t=1}^{T} d(X_t, \hat{X}_t) \le D.
\end{align}
\end{theorem}
\begin{proof}
    See Appendix \ref{app:theorem_GB}.
\end{proof}

\section{Experiments}
\label{sec:exp}
We now turn to the empirical evaluation of the proposed zero-delay online schemes on an English text compression task, and compare their performance against state-of-the-art prediction-powered compression methods \cite{valmeekam2023llmzip}.

\subsection{Setting}

We evaluate compression on English text from two sources: Shakespeare's works and Taylor Swift's song lyrics, using the datasets provided in \cite{karpathy2015charrnn} and \cite{huggingartists}, respectively.

The prediction-powered compression schemes are based on a predictor $\hat{p}_t(\cdot|Y_t)$ implemented as a \gls{llm}. Specifically, the text is first tokenized, and the distribution of the next token is modeled using the nanoGPT-2 model \cite{Karpathy2022}, which is fine-tuned on a holdout portion of the Shakespeare dataset. The nanoGPT-2 model uses a vocabulary of $|\mathcal{X}|=\rm{50257}$ tokens, which defines the alphabet of symbols to be encoded. As such, without compression, transmitting the token sequence without the use of a predictive model would require $16$ bits per symbol.

We target requirements \eqref{eq:finite_sample_dist_guarantee} expressed via: $(i)$ the outage distortion \eqref{eq:0-1distortion}, and $(ii)$ the semantic distortion measured using the normalized cosine similarity between token embeddings \cite{feng2024semantic}
\begin{align}
    \label{eq:semantic_distortion}
    d(\hat{X},X) = \frac{1}{2}\left(1 - \frac{\langle \phi(\hat{X}), \phi(X) \rangle}
    {\|\phi(\hat{X})\| \, \|\phi(X)\|}\right),
\end{align}
where $\phi(\cdot)$ denotes the embedding function mapping a token into the corresponding embedding vector produced by nanoGPT-2.

\subsection{Benchmarks}
The performance of the proposed online conformal prediction-based schemes is compared against the following benchmarks:
\begin{itemize}
    \item \textbf{LLMZip-Dropout}  \cite{valmeekam2023llmzip}, a lossless compression scheme that encodes each token $X_t$  by using an entropy coding scheme based on the model's predictive distribution $\hat{p}_t(\cdot|\hat{X}^{t-1})$. To satisfy the distortion constraint in \eqref{eq:finite_sample_dist_guarantee} for the outage distortion measure, we assume that the encoder declares an outage independently with probability $D$ at each round $t$. When an outage occurs, the transmitter and receiver agree to set $\hat{X}_t=\argmax_{x\in\mathcal{X}}\hat{p}_t(\cdot|\hat{X}^{t-1})$.
    
   \item \textbf{Block-CSC}, a variant of \gls{ocsc} designed for noiseless channels, where the encoder has access to the entire sequence to be transmitted and selects the largest threshold value $s_t$ that satisfies the distortion requirement \eqref{eq:finite_sample_dist_guarantee}  for a target distortion $D$. This scheme serves as a benchmark to compare the performance of \gls{ocsc}, which relies on online adaptation and zero-delay transmission, against a scheme that operates offline with a delay equal to the time horizon $T$, and selects the threshold value optimally in hindsight.

    \item \textbf{Block-CRDC}, a variant of \gls{ocrdc} designed for noiseless channels. In a similar manner to Block-CSC, the encoder has access to the entire source sequence as a block and selects the smallest slope parameter $s_t$ that satisfies the distortion requirement \eqref{eq:finite_sample_dist_guarantee} for a target distortion $D$. This scheme is used to compare the performance of \gls{ocrdc} against that of an offline scheme that selects the slope parameter optimally in hindsight.
    \end{itemize}
\subsection{Ideal Communication}
\begin{figure}[ht]
\centering
\includegraphics[width=0.75\linewidth]{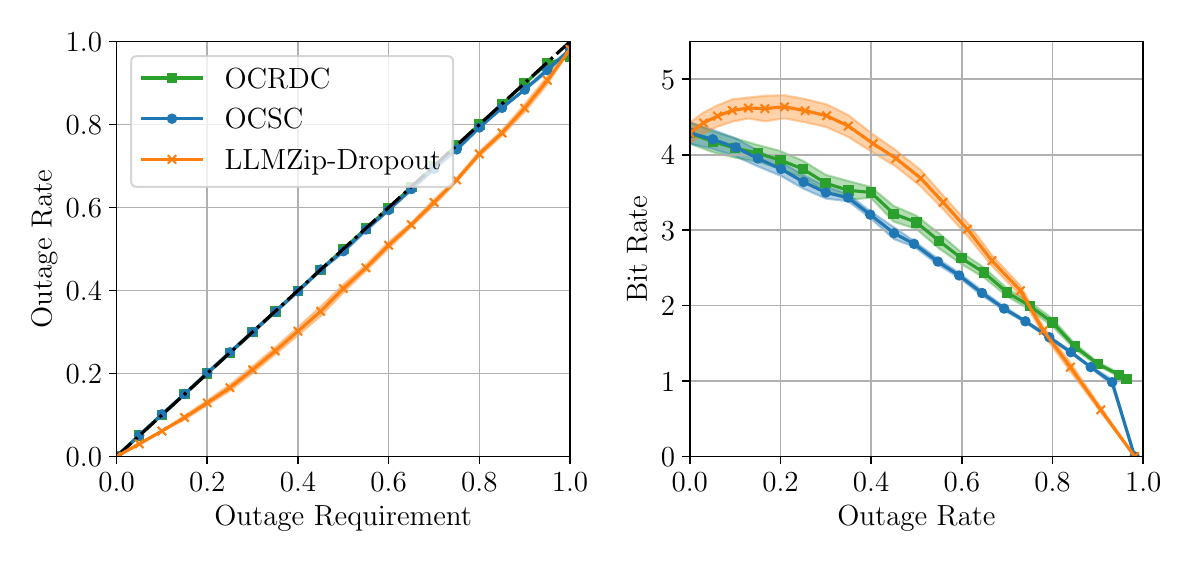}
\caption{Performance of \gls{ocrdc}, \gls{ocsc}, and LLMZip-Dropout \cite{valmeekam2023llmzip} on a sequence of 3000 tokens from the Shakespeare dataset under outage distortion requirements and error-free communication. The left panel shows the outage rate as a function of the outage requirement, while the right panel shows the bit rate as a function of the outage rate. Results are averaged over 5 runs.}
\label{fig:noiseless_outage}
\end{figure}
In this section, we evaluate the proposed schemes under ideal communication, i.e., in the absence of erasures, and under distortion requirements expressed via the outage distortion in \eqref{eq:0-1distortion}.

Fig.~\ref{fig:noiseless_outage} reports the results for compressing a sequence of $T=3000$ tokens from the Shakespeare dataset under a varying outage requirement $D$. 
The left panel of Fig.~\ref{fig:noiseless_outage} empirically validates the distortion guarantees of the schemes. 
For every target level $D$, the observed outage rate remains below $D$. 
Moreover, \gls{ocsc} and \gls{ocrdc} achieve outage rates close to the target requirement, while LLMZip-Dropout is over-conservative, maintaining substantially lower outage rates than necessary.  
The right panel shows the average bit rate achieved by \gls{ocsc}, \gls{ocrdc}, and LLMZip-Dropout as a function of the observed outage rate.
Both \gls{ocsc} and \gls{ocrdc} exhibit monotonically decreasing bit rates as the allowed outage rate increases, and they outperform LLMZip-Dropout for $D<0.75$. 
For instance, at $D=0.4$, \gls{ocsc} achieves a savings of approximately $1$ bit per token compared to LLMZip-Dropout, while \gls{ocrdc} saves about $0.8$ bits per token. 

In contrast, for target distortions up to $D=0.45$, the bit rate of LLMZip-Dropout is even higher than that of lossless transmission, i.e., $D=0$. This occurs because LLMZip-Dropout erases symbols independently of the source symbols, thereby degrading the context and the predictive distribution of the model. It is also worth noting that \gls{ocsc}, being specifically designed for outage distortion, outperforms \gls{ocrdc} for distortion levels $D>0.4$.  

Next, we examine performance under highly non-stationary data by combining text from two different sources. Specifically, we construct a sequence of $T=6000$ tokens in which the first half is sampled from Shakespeare's works and the second half from Taylor Swift's songs. Recall that the predictive model is fine-tuned on Shakespeare's data and not on Taylor Swift's songs.  

\begin{figure}[ht]
\centering
\includegraphics[width=0.8\linewidth]{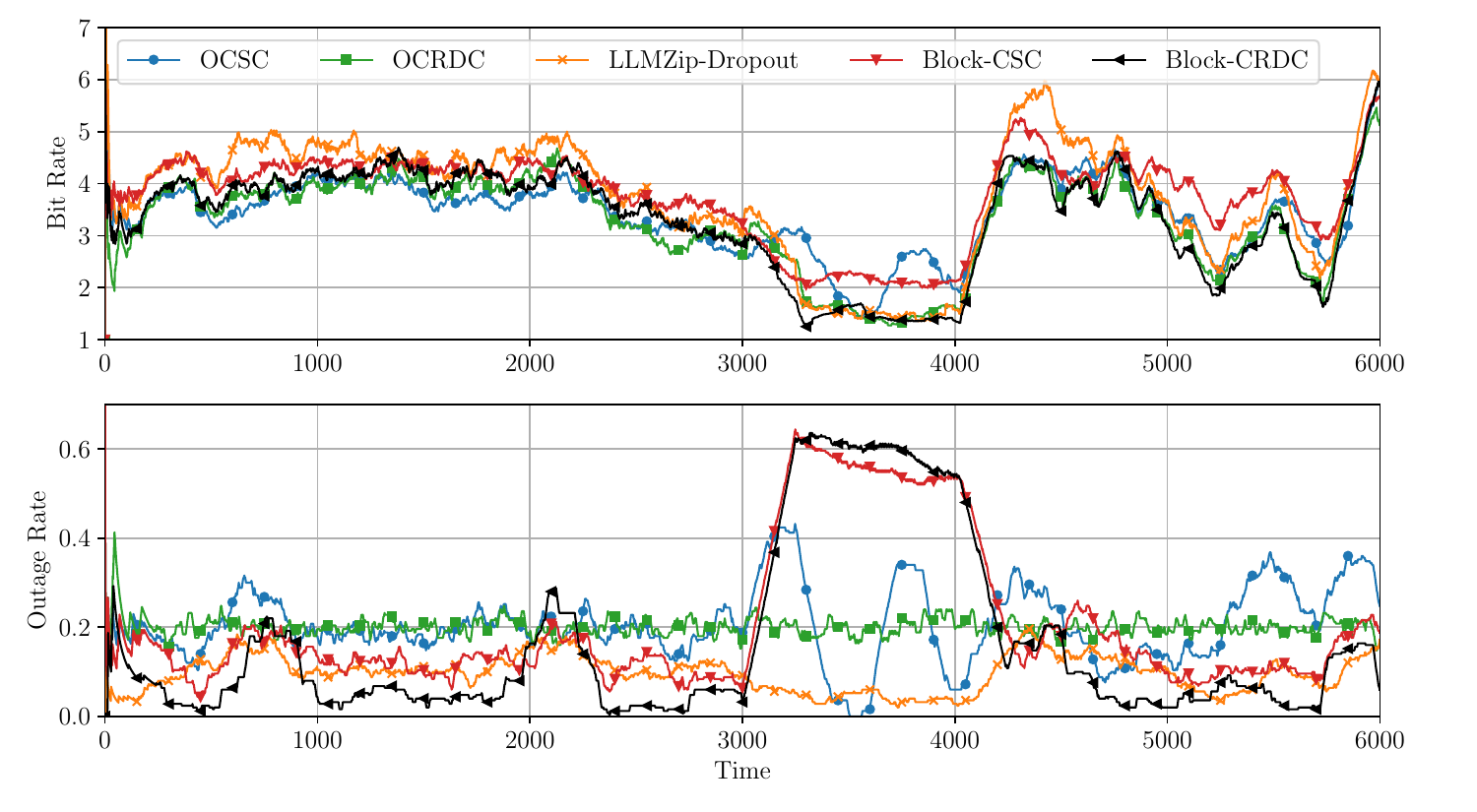}
\caption{Bit rate and outage rate of OCSC, OCRDC, LLMZip-Dropout \cite{valmeekam2023llmzip}, Block-CSC, and Block-CRDC for a sequence of $T=6000$ tokens where the first half of the tokens comes from  Shakespeare's works while the second half comes from Taylor Swift's song lyrics. }
\label{fig:non_stationary}
\end{figure}

In Fig.~\ref{fig:non_stationary}, we fix the target outage distortion at $D=0.2$, and show the bit rate and outage rate over time, averaged using a moving window of 250 time steps. In this experiment, we also include Block-CSC and Block-CRDC, for which the optimal threshold and slope parameters are determined via a grid search. All methods achieve similar bit rates, which remain relatively stable during the first half of the sequence, but become more variable in the second half.   

In terms of outage rate, all schemes satisfy the target requirement on average. However, comparing \gls{ocsc} and \gls{ocrdc} to Block-CSC and Block-CRDC, we observe that the online schemes adapt to the data source and deliver a more stable outage rate. In contrast, Block-CSC and Block-CRDC exhibit large variations in the outage rate when compressing tokens from Taylor Swift's songs. Finally, LLMZip-Dropout remains conservative, maintaining an outage rate well below the target.

\subsection{Erasure Channels}
In this section, we study compression under communication channels affected by erasures, with a target distortion requirement expressed via the semantic distortion \eqref{eq:semantic_distortion}. We consider two types of erasure channels: a memoryless erasure channel with a fixed erasure probability, and a Gilbert--Elliott channel where the erasure probability evolves over time depending on the channel's memory state. The channel parameters are chosen so that both models yield the same average erasure probability. Specifically, for the memoryless channel, we set $E_t \sim \text{Bern}(e)$ with $e=0.2$, while for the Gilbert--Elliott channel, we set the transition probabilities in \eqref{eq:gb_matrix} to $p_{\rm GB}=0.05$ and $p_{\rm BG}=0.2$, and the state-dependent erasure probabilities in \eqref{eq:gb_erasure_prob} to $e_{\rm G}=0$ and $e_{\rm B}=1$. For these parameters, the steady-state erasure probability in \eqref{eq:gb_steady_state} is $\bar{e}=0.2$, matching the erasure probability of the memoryless channel. This setup allows us to isolate and evaluate the effect of channel memory and bursty errors on the performance of the compression algorithms.

\begin{figure}[ht]
\centering
\includegraphics[width=0.75\linewidth]{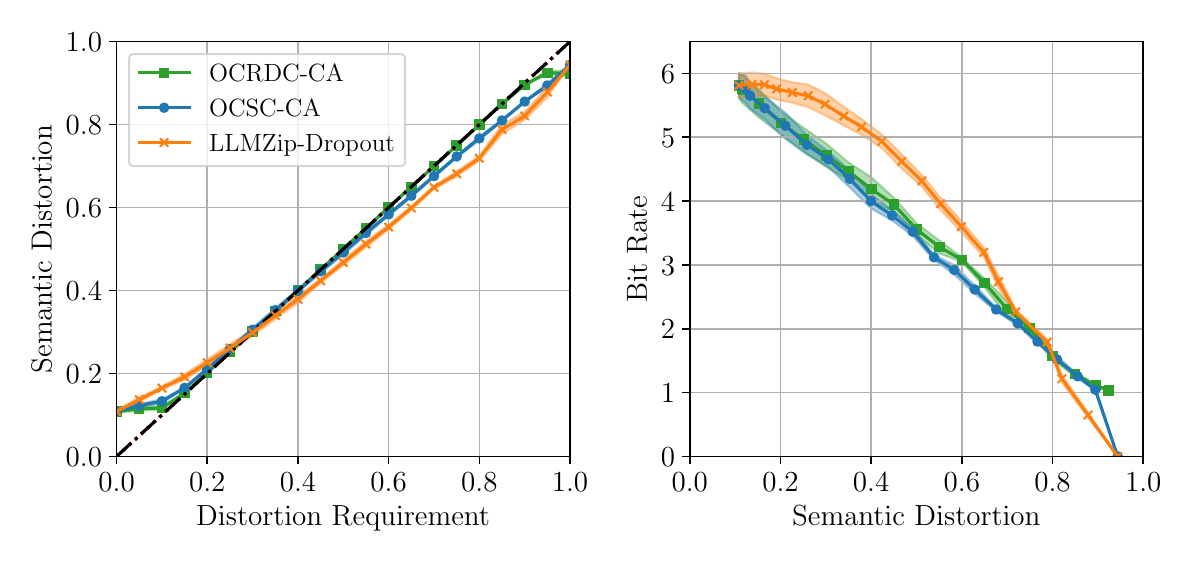}
\caption{Performance of CA-OCRDC, CA-OCSC, and LLMZip-Dropout \cite{valmeekam2023llmzip} on a sequence of 3000 tokens from the Shakespeare dataset under semantic distortion requirements and communication over a memoryless erasure channel. The left panel shows the semantic distortion as a function of the semantic distortion requirement, while the right panel shows the bit rate as a function of the semantic distortion. Results are averaged over 5 runs.}
\label{fig:erasure_fixed}
\end{figure}

\begin{figure}[ht]
\centering
\includegraphics[width=0.75\linewidth]{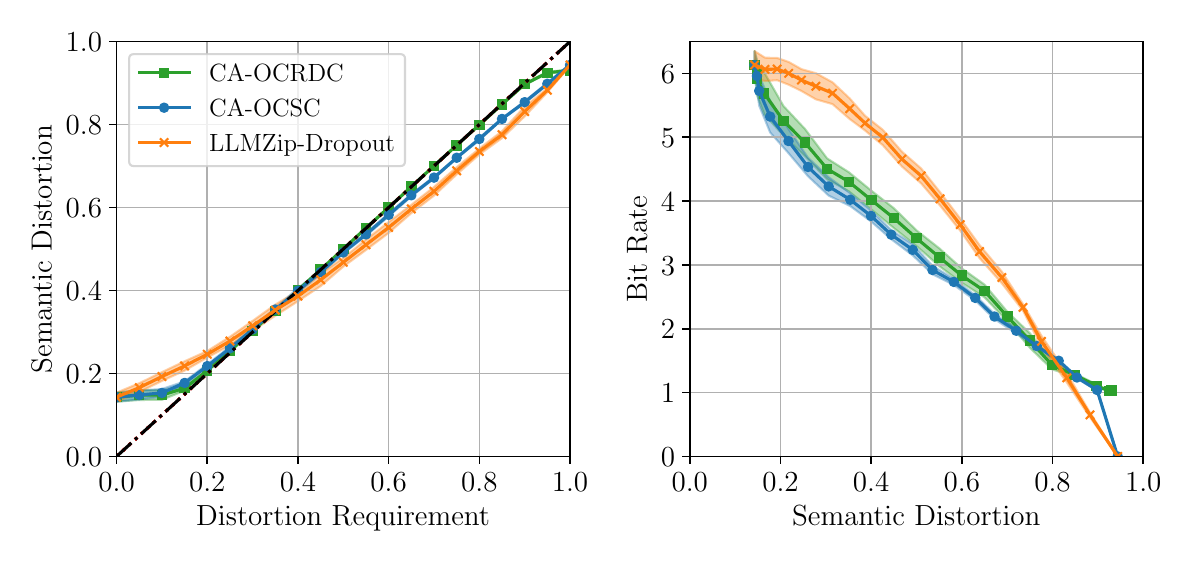}
\caption{Performance of CA-OCRDC, CA-OCSC, and LLMZip-Dropout \cite{valmeekam2023llmzip} on a sequence of 3000 tokens from the Shakespeare dataset under semantic requirements and communication over a Gilbert-Elliott erasure channel. The left panel shows the semantic distortion as a function of the semantic distortion requirement, while the right panel shows the bit rate as a function of the semantic distortion. Results are averaged over 5 runs.}
\label{fig:gb_channel}
\end{figure}

We compare the performance of the channel-aware variants of \gls{ocsc} and \gls{ocrdc}, as well as LLMZip-Dropout. 
Note that in these experiments, we consider distortion requirements in terms of semantic distortion, 
while \gls{ocsc} and LLMZip-Dropout, by design, only support requirements expressed in terms of outage distortion 
\eqref{eq:0-1distortion}. Since the semantic distortion in \eqref{eq:semantic_distortion} is between 0 and 1, the outage distortion provides a valid upper bound on the semantic distortion. Hence, OCSC and LLMZip-Dropout target a conservative requirement.

In Figs.~\ref{fig:erasure_fixed} and \ref{fig:gb_channel}, we report the average bit rate and semantic distortion 
as functions of the distortion requirement $D$, for the erasure channel and the Gilbert--Elliott channel, respectively. 
For a distortion requirement level $D \leq 0.2$, the assumptions of Theorems \ref{th:dist_guarantee_erasure_ch} and \ref{th:theorem_GB} are not satisfied, and
the compression schemes cannot meet the distortion requirement \eqref{eq:finite_sample_dist_guarantee}, delivering a semantic distortion larger than the target level $D$ due to channel-induced distortion. In contrast, for a distortion requirement level $D> 0.2$, the assumptions of Theorems \ref{th:dist_guarantee_erasure_ch} and \ref{th:theorem_GB} are satisfied, and the channel-aware schemes can compensate for the effect of channel errors. In particular, CA-OCRDC, which is designed for semantic distortion, achieves the target distortion precisely, while CA-OCSC and LLMZip-Dropout are over-conservative, since they operate under an outage distortion requirement.

Finally, when comparing performance under the two channel models, we observe that for target distortions $D < 0.2$, all schemes perform worse under the Gilbert-Elliott channel than under the memoryless erasure channel. This difference can be attributed to the nature of the errors introduced by each channel and to their impact on the accuracy of the prediction model. In the Gilbert-Elliott model, erasures occur in bursts, creating contiguous blocks of missing information that disrupt the context used by the predictive model to infer the next source symbol. In contrast, the memoryless erasure channel introduces erasures more uniformly, making it less likely for large contiguous portions of the context to be entirely lost. This highlights the impact of channel memory on the performance of prediction-powered compression based on sequence models.
\section{Conclusion}
\label{sec:conclusion} 
  We  have proposed two novel zero-delay lossy compression algorithms, \gls{ocsc} and \gls{ocrdc}, which leverage pre-trained predictors to compress arbitrary data sequences while providing finite-sample distortion guarantees under outage and general distortion measures. These distortion guarantees have been established both for ideal communication, where transmitted messages are received error-free, and for erasure channels. For the latter scenario, we have introduced channel-adaptive variants of the algorithms, supported by a novel doubly-adaptive conformal prediction rule that accounts for channel-induced distortion and provides finite-sample distortion guarantees for both deterministic and stochastic channel models. Experimental results validated our theoretical findings, showing that the proposed methods not only satisfy strict distortion requirements but also achieve superior compression efficiency compared to existing approaches.

\bibliographystyle{IEEEtran}
\bibliography{references}

\begin{thebibliography}{10}
\providecommand{\url}[1]{#1}
\csname url@samestyle\endcsname
\providecommand{\newblock}{\relax}
\providecommand{\bibinfo}[2]{#2}
\providecommand{\BIBentrySTDinterwordspacing}{\spaceskip=0pt\relax}
\providecommand{\BIBentryALTinterwordstretchfactor}{4}
\providecommand{\BIBentryALTinterwordspacing}{\spaceskip=\fontdimen2\font plus
\BIBentryALTinterwordstretchfactor\fontdimen3\font minus \fontdimen4\font\relax}
\providecommand{\BIBforeignlanguage}[2]{{%
\expandafter\ifx\csname l@#1\endcsname\relax
\typeout{** WARNING: IEEEtran.bst: No hyphenation pattern has been}%
\typeout{** loaded for the language `#1'. Using the pattern for}%
\typeout{** the default language instead.}%
\else
\language=\csname l@#1\endcsname
\fi
#2}}
\providecommand{\BIBdecl}{\relax}
\BIBdecl

\bibitem{shi2023task}
Y.~Shi, Y.~Zhou, D.~Wen, Y.~Wu, C.~Jiang, and K.~B. Letaief, ``Task-oriented communications for {6G}: Vision, principles, and technologies,'' \emph{IEEE Wireless Communications}, vol.~30, no.~3, pp. 78--85, 2023.

\bibitem{shen2024large}
Y.~Shen, J.~Shao, X.~Zhang, Z.~Lin, H.~Pan, D.~Li, J.~Zhang, and K.~B. Letaief, ``Large language models empowered autonomous edge {AI} for connected intelligence,'' \emph{IEEE Communications Magazine}, vol.~62, no.~10, pp. 140--146, 2024.

\bibitem{popovski2020semantic}
P.~Popovski, O.~Simeone, F.~Boccardi, D.~G{\"u}nd{\"u}z, and O.~Sahin, ``Semantic-effectiveness filtering and control for post-{5G} wireless connectivity,'' \emph{Journal of the Indian Institute of Science}, vol. 100, no.~2, pp. 435--443, 2020.

\bibitem{strinati2024goal}
E.~C. Strinati, P.~Di~Lorenzo, V.~Sciancalepore, A.~Aijaz, M.~Kountouris, D.~G{\"u}nd{\"u}z, P.~Popovski, M.~Sana, P.~A. Stavrou, B.~Soret \emph{et~al.}, ``Goal-oriented and semantic communication in {6G} {AI}-native networks: The {6G-GOALS} approach,'' in \emph{2024 Joint European Conference on Networks and Communications \& 6G Summit (EuCNC/6G Summit)}.\hskip 1em plus 0.5em minus 0.4em\relax IEEE, 2024, pp. 1--6.

\bibitem{guo2024distributed}
J.~Guo, H.~Chen, B.~Song, Y.~Chi, C.~Yuen, F.~R. Yu, G.~Y. Li, and D.~Niyato, ``Distributed task-oriented communication networks with multimodal semantic relay and edge intelligence,'' \emph{IEEE Communications Magazine}, vol.~62, no.~6, pp. 82--89, 2024.

\bibitem{getu2023making}
T.~M. Getu, G.~Kaddoum, and M.~Bennis, ``Making sense of meaning: A survey on metrics for semantic and goal-oriented communication,'' \emph{IEEE Access}, vol.~11, pp. 45\,456--45\,492, 2023.

\bibitem{pourkabirian2024vision}
A.~Pourkabirian, M.~S. Kordafshari, A.~Jindal, and M.~H. Anisi, ``A vision of {6G URLLC}: Physical-layer technologies and enablers,'' \emph{IEEE Communications Standards Magazine}, vol.~8, no.~2, pp. 20--27, 2024.

\bibitem{shannon1959coding}
C.~E. Shannon, ``Coding theorems for a discrete source with a fidelity criterion,'' in \emph{IRE National Convention Record, 1959}, vol.~4, 1959, pp. 142--163.

\bibitem{berger2003rate}
T.~Berger, ``Rate-distortion theory,'' \emph{Wiley Encyclopedia of Telecommunications}, 2003.

\bibitem{davisson1973universal}
L.~Davisson, ``Universal noiseless coding,'' \emph{IEEE Transactions on Information Theory}, vol.~19, no.~6, pp. 783--795, 1973.

\bibitem{ziv2003coding}
J.~Ziv, ``Coding theorems for individual sequences,'' \emph{IEEE Transactions on Information Theory}, vol.~24, no.~4, pp. 405--412, 1978.

\bibitem{jalali2008rate}
S.~Jalali and T.~Weissman, ``Rate-distortion via {M}arkov chain {M}onte {C}arlo,'' in \emph{2008 IEEE International Symposium on Information Theory (ISIT)}.\hskip 1em plus 0.5em minus 0.4em\relax IEEE, 2008, pp. 852--856.

\bibitem{jalali2011lossy}
S.~Jalali, A.~Montanari, and T.~Weissman, ``Lossy compression of discrete sources via the {V}iterbi algorithm,'' \emph{IEEE Transactions on Information Theory}, vol.~58, no.~4, pp. 2475--2489, 2011.

\bibitem{yang1997fixed}
E.-h. Yang, Z.~Zhang, and T.~Berger, ``Fixed-slope universal lossy data compression,'' \emph{IEEE Transactions on Information Theory}, vol.~43, no.~5, pp. 1465--1476, 1997.

\bibitem{linder2001zero}
T.~Linder and G.~Lugosi, ``A zero-delay sequential scheme for lossy coding of individual sequences,'' \emph{IEEE Transactions on Information Theory}, vol.~47, no.~6, pp. 2533--2538, Sep. 2001.

\bibitem{gyorgy2004efficient}
A.~Gyorgy, T.~Linder, and G.~Lugosi, ``Efficient adaptive algorithms and minimax bounds for zero-delay lossy source coding,'' \emph{IEEE Transactions on Signal Processing}, vol.~52, no.~8, pp. 2337--2347, 2004.

\bibitem{matloub2006universal}
S.~Matloub and T.~Weissman, ``Universal zero-delay joint source--channel coding,'' \emph{IEEE Transactions on Information Theory}, vol.~52, no.~12, pp. 5240--5250, 2006.

\bibitem{shalev2012online}
S.~Shalev-Shwartz \emph{et~al.}, ``Online learning and online convex optimization,'' \emph{Foundations and Trends in Machine Learning}, vol.~4, no.~2, pp. 107--194, 2012.

\bibitem{gibbs2021adaptive}
I.~Gibbs and E.~Candes, ``Adaptive conformal inference under distribution shift,'' \emph{Proceedings of the Advances in Neural Information Processing Systems (NeurIPS)}, vol.~34, pp. 1660--1672, Dec. 2021.

\bibitem{feldman2023achieving}
\BIBentryALTinterwordspacing
S.~Feldman, L.~Ringel, S.~Bates, and Y.~Romano, ``Achieving risk control in online learning settings,'' \emph{Transactions on Machine Learning Research}, 2023. [Online]. Available: \url{https://openreview.net/forum?id=5Y04GWvoJu}
\BIBentrySTDinterwordspacing

\bibitem{angelopoulos2024online}
A.~N. Angelopoulos, R.~Barber, and S.~Bates, ``Online conformal prediction with decaying step sizes,'' in \emph{Proceedings of the International Conference on Machine Learning (ICML)}, Vienna, Austria, Jul. 2024, pp. 1616--1630.

\bibitem{angelopoulos2025gradient}
A.~N. Angelopoulos, M.~I. Jordan, and R.~J. Tibshirani, ``Gradient equilibrium in online learning: Theory and applications,'' \emph{arXiv preprint arXiv:2501.08330}, 2025.

\bibitem{ganesan2025online}
U.~K. Ganesan, G.~Durisi, M.~Zecchin, P.~Popovski, and O.~Simeone, ``Online conformal compression for zero-delay communication with distortion guarantees,'' \emph{arXiv preprint arXiv:2503.08340}, 2025.

\bibitem{dai2022nonlinear}
J.~Dai, S.~Wang, K.~Tan, Z.~Si, X.~Qin, K.~Niu, and P.~Zhang, ``Nonlinear transform source-channel coding for semantic communications,'' \emph{IEEE Journal on Selected Areas in Communications}, vol.~40, no.~8, pp. 2300--2316, 2022.

\bibitem{Simeone_2025}
O.~Simeone, \emph{Classical and Quantum Information Theory: Uncertainty, Information, and Correlation}.\hskip 1em plus 0.5em minus 0.4em\relax Cambridge University Press, 2025.

\bibitem{neely2010stochastic}
M.~Neely, \emph{Stochastic network optimization with application to communication and queueing systems}.\hskip 1em plus 0.5em minus 0.4em\relax Morgan \& Claypool Publishers, 2010.

\bibitem{chang1994stability}
C.-S. Chang, ``Stability, queue length, and delay of deterministic and stochastic queueing networks,'' \emph{IEEE Transactions on Automatic Control}, vol.~39, no.~5, pp. 913--931, 1994.

\bibitem{gilbert1960capacity}
E.~Gilbert, ``Capacity of a burst-noise channel,'' \emph{The Bell System Technical Journal}, vol.~39, no.~5, pp. 1253--1265, 1960.

\bibitem{elliott1963estimates}
E.~O. Elliott, ``Estimates of error rates for codes on burst-noise channels,'' \emph{The Bell System Technical Journal}, vol.~42, no.~5, pp. 1977--1997, 1963.

\bibitem{paulin2015concentration}
D.~Paulin, ``Concentration inequalities for {M}arkov chains by {M}arton couplings and spectral methods,'' \emph{Electronic Journal of Probability}, vol.~20, no.~79, pp. 1--32, 2015.

\bibitem{valmeekam2023llmzip}
C.~S.~K. Valmeekam, K.~Narayanan, D.~Kalathil, J.-F. Chamberland, and S.~Shakkottai, ``{LLMzip}: {L}ossless text compression using large language models,'' \emph{arXiv:2306.04050}, 2023.

\bibitem{karpathy2015charrnn}
A.~Karpathy, \url{https://raw.githubusercontent.com/karpathy/char-rnn/master/data/tinyshakespeare/input.txt}, 2015.

\bibitem{huggingartists}
A.~Korshuk, \url{https://huggingface.co/datasets/huggingartists/taylor-swift}, 2021.

\bibitem{Karpathy2022}
\BIBentryALTinterwordspacing
A.~Karpathy, ``nano{GPT}: The simplest, fastest repository for training/finetuning medium-sized {GPT}s,'' 2022, accessed: 2025-09-25. [Online]. Available: \url{https://github.com/karpathy/nanoGPT}
\BIBentrySTDinterwordspacing

\bibitem{feng2024semantic}
Y.~Feng, ``Semantic textual similarity analysis of clinical text in the era of {LLM},'' in \emph{2024 IEEE Conference on Artificial Intelligence (CAI)}.\hskip 1em plus 0.5em minus 0.4em\relax IEEE, 2024, pp. 1284--1289.

\bibitem{lezaud1998chernoff}
P.~Lezaud, ``Chernoff-type bound for finite {M}arkov chains,'' \emph{Annals of Applied Probability}, pp. 849--867, 1998.

\bibitem{fan2021hoeffding}
J.~Fan, B.~Jiang, and Q.~Sun, ``Hoeffding's inequality for general {M}arkov chains and its applications to statistical learning,'' \emph{Journal of Machine Learning Research}, vol.~22, no. 139, pp. 1--35, 2021.

\end{thebibliography}

\appendix
\section{Proofs}
\subsection{Proof of Theorem \ref{th:occ_guarantee_noiseless}}
\label{app:occ_guarantee_noiseless}

By iterating the update rule~\eqref{eq:update_rule_occ_noiseless} we obtain
\begin{align}
    \frac{1}{T}\sum_{t=1}^T \mathds{1}\{X_t\notin\mathcal{X}_t\}
    = D + \frac{\lambda_0-\lambda_T}{\eta T}.
\end{align}
To conclude, it suffices to show that $\lambda_t\geq -\eta(1-D)$ for all $t$.  
Assume, for contradiction, that $\lambda_{t+1}<-\eta(1-D)$ for the first time at $t+1$. Then $\lambda_t = \lambda_{t+1}+\eta\big(\mathds{1}\{X_t\notin\mathcal{X}_t\}-D\big) < 0$ and $s_t=0$ due to (\ref{eq:threshold_level}), which implies that $X_t=\hat X_t$. This forces the update at $t+1$ to increase $\lambda_{t+1}$, contradicting $\lambda_{t+1}<-\eta(1-D)$.  
Therefore $\lambda_t\geq -\eta(1-D)$ for all $t$. Since $X_t\neq \hat X_t \implies X_t\notin\mathcal{X}_t$, it holds 
\begin{align}
    \frac{1}{T}\sum_{t=1}^T \mathds{1}\{X_t\neq \hat X_t\}
    \le \frac{1}{T}\sum_{t=1}^T \mathds{1}\{X_t\notin\mathcal{X}_t\}\leq  D + \frac{\lambda_0+\eta(1-D)}{\eta T}.
\end{align}
\subsection{Proof of Lemma \ref{lemma:bounded_lambda}}
\label{proof:bounded_lambda}
   Assume, for contradiction, that $\lambda_{t+1}>L/D+\eta(D_{\max}-D)$ for the first time at $t+1$. 
Then from the update rule~\eqref{eq:update_rule_noiseless}, it follows that $\lambda_t = \lambda_{t+1}-\eta\big(d(X_t,\tilde X_t)-D\big) > L/D$. Since $\tilde X_t$ solves~\eqref{eq:rd_obj}, we have
\begin{align}
    \lambda_t d(X_t,\tilde X_t) 
    &\le -\log \hat p_t(\tilde X_t) + \lambda_t d(X_t,\tilde X_t) \le -\log \hat p_t(X_t) + \lambda_t d(X_t,X_t) = -\log \hat p_t(X_t),
\end{align}
hence
\begin{align}
    d(X_t,\tilde X_t) 
    \le \frac{-\log \hat p_t(X_t)}{\lambda_t} < D,
\end{align}
where the last inequality uses Assumption~\ref{ass:max_cost}. Being the distortion below $D$, the update decreases $\lambda_{t+1}$, contradicting $\lambda_{t+1}>L/D+\eta(D_{\max}-D)$. Therefore $\lambda_t \le L/D+\eta(D_{\max}-D)$ for all $t$.
\subsection{Proof of Theorem \ref{th:distortion_noiseless}}
\label{proof:distortion_noiseless}

Telescoping the update rule \eqref{eq:update_rule_noiseless} over the $T$ time steps, and invoking Lemma  \ref{lemma:bounded_lambda} it holds
\begin{align}
    \frac{1}{T}\sum^T_{t=1}d(\hat{X}_t,X_t)=D+\frac{\lambda_T-\lambda_0}{\eta T}\leq D+\frac{L/D+\eta(D_{\rm max}-D)-\lambda_0}{\eta T}.
\end{align}

\subsection{Proof of Lemma \ref{lemma:bounded_lambda_timevarying}}
\label{app:bounded_lambda_timevarying}

\textbf{CA-OCSC.} Suppose, for contradiction, that $\lambda_{t+1}<-\eta(1-\epsilon)$ for the first time at $t+1$.  From the update rule~\eqref{eq:update_rule_noisy}, it follows
\begin{align}
    \lambda_t 
    &= \lambda_{t+1}+\eta\big((1-E_t)\mathds{1}\{X_t\notin \mathcal X_t\}-D+\delta^{\rm tgt}_t\big) < 0, \label{eq:lambda_ocsc}
\end{align}
since $\delta^{\rm tgt}_t\leq D-\epsilon$. Thus $s_t=0$ and $X_t\in\mathcal X_t$, which forces the update at $t+1$ to increase $\lambda_{t+1}$, contradicting $\lambda_{t+1}<-\eta(1-\epsilon)$. Hence $\lambda_t\geq -\eta(1-\epsilon)$ for all $t\geq 1$.

\textbf{CA-OCRDC.} Suppose, for contradiction, that $\lambda_{t+1}>L/\epsilon+\eta(D_{\max}-\epsilon)$ for the first time at $t+1$. From the update rule~\eqref{eq:update_rule_noisy}, it follows
\begin{align}
    \lambda_t 
    &= \lambda_{t+1}-\eta\big(d(X_t,\tilde X_t)-D+\delta^{\rm tgt}_t\big)> L/\epsilon. \label{eq:lambda_ocrdc}
\end{align}
Since $\tilde X_t$ solves~\eqref{eq:rd_obj},
\begin{align}
    \lambda_t d(X_t,\tilde X_t) 
    &\le -\log \hat p_t(\tilde X_t) + \lambda_t d(X_t,\tilde X_t) \le -\log \hat p_t(X_t) + \lambda_t d(X_t,X_t) = -\log \hat p_t(X_t),
\end{align}
which implies
\begin{align}
    d(X_t,\tilde X_t) 
    &\le \frac{-\log \hat p_t(X_t)}{\lambda_t} < \epsilon \le D-\delta^{\rm tgt}_t, \label{eq:dist_bound}
\end{align}
by Assumption~\ref{ass:max_cost_2}. Thus the distortion is below the adjusted target, so the update decreases $\lambda_{t+1}$, again contradicting the hypothesis. Therefore $\lambda_t \le L/\epsilon+\eta(D_{\max}-\epsilon)$  for all $t\geq 1$.

\subsection{Proof of Proposition \ref{prop:distortion_queue}}
\label{app:distortion_queue}

\textbf{CA-OCSC.} The distortion of the received sequence $\hat{X}^T$ can be upper bounded as 
\begin{align}   
\frac{1}{T}\sum^{T}_{t=1} d( X_t,\hat{X}_t)=\frac{1}{T}\sum^{T}_{t=1} \mathds{1}\{X_t\notin \mathcal{X}_t\}
&=\frac{1}{T}\sum^{T}_{t=1} (1-E_t) \mathds{1}\{X_t\notin \mathcal{X}_t\}+\frac{1}{T}\sum^{T}_{t=1} E_t \mathds{1}\{X_t\notin \mathcal{X}_t\} \nonumber\\
&\leq \frac{1}{T}\sum^{T}_{t=1} (1-E_t) \mathds{1}\{X_t\notin \mathcal{X}_t\} +\frac{1}{T}\sum^{T}_{t=1} \tilde\delta_t^{\rm ch}\label{eq:actual_distortion}
\end{align}
At the same time, telescoping the update rule \eqref{eq:update_rule_noisy} and dividing by the horizon $T$ yields 
\begin{align}
\label{eq:telescoped_adjusted}
\frac{1}{T} \sum^T_{t=1}(1-E_t) \mathds{1}\{X_t \notin \mathcal{X}_t\} &= D + \frac{\lambda_0 - \lambda_T}{\eta T} - \frac{1}{T} \sum^T_{t=1} \delta^{\rm tgt}_t \leq D + \frac{\eta (1 - \epsilon) - \lambda_0}{\eta T} - \frac{1}{T} \sum^T_{t=1} \delta^{\rm tgt}_t,
\end{align} 
where the last inequality follows from Lemma \ref{lemma:bounded_lambda_timevarying}. Plugging the inequality \eqref{eq:telescoped_adjusted} in \eqref{eq:actual_distortion} yields the desired result.

\textbf{CA-OCRDC}. The distortion of the received sequence $\hat{X}^T$ can be equivalently expressed as the cumulative distortion of the transmitted sequence $\tilde{X}^T$ plus the upper bound on the cumulative channel-induced distortion
\begin{align}   
\label{eq:actual_distortion_2}
\frac{1}{T}\sum^{T}_{t=1} d(X_t, \hat{X}_t) &= \frac{1}{T}\sum^T_{t=1} \tilde\delta_t^{\rm ch} + \frac{1}{T}\sum^T_{t=1} d(X_t, \tilde{X}_t).
\end{align}
Telescoping the update rule \eqref{eq:update_rule_noisy} for CA-OCRDC, and dividing by the horizon $T$ we obtain
\begin{align}
\label{eq:telescoped_adjusted_2}
\frac{1}{T} \sum^T_{t=1} d(X_t, \tilde{X}_t) &= D + \frac{\lambda_T - \lambda_0}{\eta T} - \frac{1}{T} \sum^T_{t=1} \delta^{\rm tgt}_t\leq D + \frac{L/\epsilon + \eta (D_{\rm max} - \epsilon) - \lambda_0}{\eta T} - \frac{1}{T} \sum^T_{t=1} \delta^{\rm tgt}_t,
\end{align} 
where the last inequality follows from Lemma \ref{lemma:bounded_lambda_timevarying}. Plugging the term \eqref{eq:telescoped_adjusted_2} in \eqref{eq:actual_distortion_2} we obtain the desired result.

\subsection{Proof of Theorem \ref{th:general_dist_guarantee_noisy}}
\label{app:general_dist_guarantee_noisy}

We show that under the erasure condition \eqref{eq:distortion_cond}, the queue size $Q_t$ is bounded. Specifically, for any interval $[t_1, t_2)$ during which $Q_t > D - \epsilon$ for all $t \in [t_1, t_2)$, with $Q_{t_1-1} < D - \epsilon$ and $Q_{t_2} < D - \epsilon$, we show that the interval length $\tau = t_2 - t_1$ is bounded by a constant.

 Note that if, at time $t$, we have $Q_t \geq D - \epsilon$, then it follows from \eqref{eq:adjustment_policy} that the service rate at the next step is $\delta^{\text{tgt}}_{t+1} = D - \epsilon$, whereas if $Q_t < D - \epsilon$, it holds $\delta^{\text{tgt}}_{t+1} = Q_t$ and the queue size at time $t+1$ equals $Q_{t+1} = \delta^{\rm ch}_{t+1}$. The queue size $Q_{t_2}$ can then be bounded as 
\begin{align}
    Q_{t_2}=\sum^{t_2}_{t=t_1}\tilde\delta^{\rm ch}_t - \delta^{\rm tgt}_t&=\sum^{t_2}_{t=t_1}\tilde\delta^{\rm ch}_t - (\tau-1)(D - \epsilon) \\
    &\leq D_{\rm max}\sum^{t_2}_{t=t_1}E_t - (\tau-1)(D - \epsilon) \label{eq:ineq_1_th3}\\
    &\leq (AD_{\rm max} - D + \epsilon)\tau+ D_{\rm max}\psi(\tau)+(D - \epsilon), \label{eq:ineq_2_th3}
\end{align}
 where the inequality \eqref{eq:ineq_1_th3} follows because the channel-induced distortion is $0$ in case of error-free transmission and is bounded by $D_{\rm max}$ in case of erasure, and the second inequality follows from Assumption \ref{ass:bounded_channel_distortion}.
Since $A < (D - \epsilon)/D_{\rm max}$ by Assumption \ref{ass:bounded_channel_distortion}, the coefficient $(A D_{\rm max} - D + \epsilon)$ is negative, which ensures that, for sufficiently large $\tau$, the queue will eventually fall below $D - \epsilon$; more precisely, the maximum duration of intervals during which the queue exceeds $D - \epsilon$ is uniformly bounded by $\tau_{\rm max}$, as defined in \eqref{eq:tau_max}.  
Being the maximum increment of the queue at each time instant in these intervals is bounded by $D_{\rm max}-D+\epsilon$, the queue size satisfies
\begin{align}
    \label{eq:bound_queue_th3}
    \sup_{t} Q_t < \tau_{\rm max}(D_{\rm max}-D+\epsilon)+D-\epsilon.
\end{align}
We obtain the desired result by substituting the queue bound \eqref{eq:bound_queue_th3} in \eqref{eq:intermediate_bound}.

\subsection{Proof of Theorem \ref{th:dist_guarantee_erasure_ch}}
\label{app:dist_guarantee_erasure_ch}

    The size of the residual distortion queue at time $T$ can be upper bounded as \cite{neely2010stochastic}
    \begin{align}
        Q_T=\max_{0\leq s\leq T}\left\{\sum^T_{t=T-s}\tilde\delta^{\rm ch}_t-s(D-\epsilon)\right\}\leq \max_{0\leq s\leq T}\left\{\sum^T_{t=T-s}E_tD_{\rm max}-s(D-\epsilon)\right\},
    \end{align}
    where the inequality follows from the fact that the channel-induced distortion is $0$ in case of error-free transmission and is bounded by $D_{\rm max}$ in case of erasure.
    Using a union bound argument, the probability that the queue at time $T$ exceeds a constant $r>0$ can be bounded as
    \begin{align}
        \label{eq:prob_bound_queue_sum}
        \Pr\left[\max_{0\leq s\leq T}\left\{\sum^T_{t=T-s}E_tD_{\rm max}-s(D-\epsilon)\right\}>r\right]\leq \sum^T_{s=0}\Pr\left[\sum^T_{t=T-s}E_t>\frac{r+s(D-\epsilon)}{D_{\rm max}}\right].
    \end{align}
    Chernoff bound applied to each term of \eqref{eq:prob_bound_queue_sum} yields for any $\theta>0$,
     \begin{align}
       \Pr\left[\sum^T_{t=T-s}E_t>\frac{r+s(D-\epsilon)}{D_{\rm max}}\right]&\leq e^{-\theta \frac{r+s(D-\epsilon)}{D_{\rm max}}}\mathbb{E}\left[e^{\theta\sum^T_{t=T-s}E_t}\right]\nonumber\\
       &=e^{-\theta \frac{r+s(D-\epsilon)}{D_{\rm max}}}\prod^{T}_{t=T-s}\mathbb{E}\left[e^{\theta E_t}\right]\nonumber\\
       &\leq e^{-\theta \frac{r+s(D-\epsilon)}{D_{\rm max}}}e^{(e^{\theta}-1)\sum^T_{t=T-s}e_t}\nonumber\\
       &=e^{-\theta \frac{s(D-\epsilon)}{D_{\rm max}}+(e^{\theta}-1)\sum^T_{t=T-s}e_t}e^{-\theta\frac{r}{D_{\rm max}}},\label{eq:intermediate_th3}
    \end{align}
    where in the last inequality we used that $1+y\leq e^{y}$.
    From Assumption \ref{ass:bounded_erasure_mep}, we obtain
      \begin{align}
       \Pr\left[\sum^T_{t=T-s}E_t>\frac{r+s(D-\epsilon)}{D_{\rm max}}\right]&\leq e^{-\theta \frac{s(D-\epsilon)}{D_{\rm max}}+(e^{\theta}-1)(A s +\psi(s))}e^{-\theta\frac{r}{D_{\rm max}}}\nonumber\\
       &=e^{s \left(-\theta \frac{(D-\epsilon)}{D_{\rm max}}+(e^{\theta}-1)A  \right)}e^{-\theta\frac{r}{D_{\rm max}}+(e^{\theta}-1)\psi(s)}
    \end{align}
    Consider the function 
    \begin{align}
        f(\theta)=-\theta \frac{(D-\epsilon)}{D_{\rm max}}+(e^{\theta}-1)A.
    \end{align}
    We want to show that there exists $\theta^*>0$ such that $f(\theta^*)<0$. Note that $f(0)=0$ and its derivative 
    \begin{align}
        f'(\theta)=- \frac{(D-\epsilon)}{D_{\rm max}}+e^{\theta}A
    \end{align}
    is such that $f'(0)=A- \frac{(D-\epsilon)}{D_{\rm max}}<0$ since, by assumption $A<\frac{(D-\epsilon)}{D_{\rm max}}$. It follows that such $\theta^*$ exists. It then follows 
    \begin{align}
      \label{eq:temp_bound_1}
      \sum^T_{s=0} \Pr\left[\sum^T_{t=T-s}E_t>\frac{r+s(D-\epsilon)}{D_{\rm max}}\right]&\leq e^{-\theta^*\frac{r}{D_{\rm max}}+(e^{\theta^*}-1)\psi(T)}\sum^T_{s=0} e^{s f(\theta^*)}.
    \end{align}
    Given that $f(\theta^*)<0$, then last term in \eqref{eq:temp_bound_1} is a geometric series with ratio $e^{f(\theta^*)}<1$ and
    \begin{align}
        \sum^T_{s=0} e^{s f(\theta^*)} \leq  \sum^{\infty}_{s=0} e^{s f(\theta^*)}\leq \frac{1}{1-e^{f(\theta^*)}}.  
    \end{align}  
    We obtain the final bound
    \begin{align}
        \Pr[Q_T>r]\leq \frac{e^{-\theta^*\frac{r}{D_{\rm max}}+(e^{\theta^*}-1)\psi(T)}}{1-e^{f(\theta^*)}}, 
    \end{align}
    which can be equivalently stated as
     \begin{align}
        \label{eq:bound_queue_prob_th4}
        \Pr\left[Q_T>\frac{D_{\rm max}}{\theta^*}\left((e^{\theta^*}-1)\psi(T)+\log\left(\frac{1}{\delta_T\left(1-e^{f(\theta^*)}\right)}\right)\right)\right]\leq \delta_T. 
    \end{align}
    Substituting \eqref{eq:bound_queue_prob_th4} into \eqref{eq:intermediate_bound}, we conclude that with probability $1-\delta_T$
    \begin{align}
     \frac{1}{T} \sum_{t=1}^{T} d(X_t, \hat{X}_t) \leq D + \frac{K}{T\eta}+\frac{D_{\rm max}}{T\theta^*}\left((e^{\theta^*}-1)\psi(T)+\log\left(\frac{1}{\delta_T\left(1-e^{f(\theta^*)}\right)}\right)\right),
    \end{align}
    where $K$ is defined as \eqref{eq:definition_K}.
    
    Setting $\delta_T=1/T^2$, the sequence of probability values satisfy $\sum^{\infty}_{T=1}\delta_T<\infty$, so the Borel-Cantelli lemma implies that, almost surely,
    \begin{align}
         \limsup_{T\to\infty}\frac{1}{T} \sum_{t=1}^{T} d(X_t, \hat{X}_t) \leq &D.
    \end{align}

\subsection{Proof of Lemma \ref{eq:lemma_GB}}
\label{app:lemma_GB}
Define the function $g(\cdot)$ that maps the state of the Markov chain $Z_t\in\{\mathrm{G},\mathrm{B}\}$ to the corresponding erasure probability $e_t\in[0,1]$. For any time index $k\in\{1,\dots,T\}$ and any state distribution $Z_k\sim P_q$, Proposition 3.10 of \cite{paulin2015concentration} allows us to bound the deviation term as 
\begin{align}
    \Pr\left[\sum_{t=k}^{k+\tau} g(Z_t)-\tau\bar{e}>\epsilon \,\Big|\, Z_k\sim P_q\right]
    &\leq \mathbb{E}_{Z\sim P_{q}}\left[\frac{P_{q}(Z)}{\pi(Z)}\right]
    \Pr\left[\sum_{t=k}^{k+\tau} g(Z_t)-\tau\bar{e}>\epsilon \,\Big|\, Z_k\sim \pi\right]^{1/2}\nonumber\\
    &\leq \big(\min\{\pi(\mathrm{G}),\pi(\mathrm{B})\}\big)^{-1}
    \Pr\left[\sum_{t=k}^{k+\tau} g(Z_t)-\tau\bar{e}>\epsilon \,\Big|\, Z_k\sim \pi\right]^{1/2}.
    \label{eq:initial_proof_lemma}
\end{align}

The probability on the right-hand side of \eqref{eq:initial_proof_lemma} can be bounded using a Hoeffding-type inequality for Markov chains \cite{lezaud1998chernoff,fan2021hoeffding}
\begin{align}
    \Pr\left[\sum_{t=k}^{k+\tau} g(Z_t)-\tau\bar{e}>\epsilon \,\Big|\, Z_k\sim \pi\right]
    \leq \exp\!\left(-\frac{1-\gamma}{1+\gamma}\frac{2\epsilon^2}{\tau}\right),
\end{align}
where $\gamma$ denotes the spectral gap of the Markov chain. Setting 
\begin{align}
\epsilon= \sqrt{ \frac{\tau}{2}\frac{1+\gamma}{1-\gamma} \log\left( \frac{T}{\delta_\tau} \right) },
\qquad 
\delta_\tau= \big(\min\{\pi(\mathrm{G}),\pi(\mathrm{B})\}\big)^2\left(\frac{6\delta}{\pi^2\tau^2}\right)^2,
\end{align}
we can apply a union bound to control the deviation simultaneously for all $k\in\{1,\dots,T\}$:
\begin{align}
    \label{eq:bound_queue_th5}
    \Pr\left[\max_{1\leq k\leq T}\sum_{t=k}^{k+\tau} g(Z_t)-\tau\bar{e}>\epsilon \,\Big|\, Z_k\sim \pi\right]
    &\leq \sum_{k=1}^T \Pr\left[\sum_{t=k}^{k+\tau} g(Z_t)-\tau\bar{e}>\epsilon \,\Big|\, Z_k\sim \pi\right]\leq \delta_\tau.
\end{align}

Combining \eqref{eq:bound_queue_th5} with \eqref{eq:initial_proof_lemma} and using a union bound, we obtain the following bound that holds uniformly over $\tau\geq 1$
\begin{align}
    \Pr\left[\exists \tau\in\mathbb{N}:\sum_{t=k}^{k+\tau} g(Z_t)>\bar{e}\tau+
    \sqrt{ \tau\frac{1+\gamma}{1-\gamma} 
    \log\left( \frac{\sqrt{T}\pi^2\tau^2}{\min\{\pi(\mathrm{G}),\pi(\mathrm{B})\}  6\delta} \right) }
    \,\Bigg|\, Z_k\sim P_q\right]
    &\leq \sum_{\tau=1}^{\infty}\frac{6\delta}{\pi^2\tau^2}\nonumber\\
    &= \delta,
\end{align}
where the last equality follows from the property of the Basel series.
\subsection{Proof of Theorem \ref{th:theorem_GB}}
\label{app:theorem_GB}
From Lemma~\ref{eq:lemma_GB}, with probability $1-\delta_T/2$, it holds 
\begin{align}
\sum_{t=k}^{k+\tau} e_t \leq \bar{e} \, \tau +
     \sqrt{ \tau\frac{1+\gamma}{1-\gamma} 
    \log\left( \frac{\sqrt{T}\pi^2\tau^2}{\min\{\pi(\mathrm{G}),\pi(\mathrm{B})\}  3\delta_T} \right) }.
\end{align}
Following the same step as in the proof of Theorem \ref{th:dist_guarantee_erasure_ch} and applying the inequality above to \eqref{eq:intermediate_th3}, we conclude that, with probability $1-\delta_T/2$,  
\begin{align}
     \frac{1}{T} \sum_{t=1}^{T}& d(X_t, \hat{X}_t) \leq D + \frac{K}{T\eta} \nonumber\\
     &+\frac{D_{\rm max}}{T\theta^*}\left((e^{\theta^*}-1)\sqrt{ T\frac{1+\gamma}{1-\gamma} 
    \log\left( \frac{\sqrt{T}\pi^2T^2}{\min\{\pi(\mathrm{G}),\pi(\mathrm{B})\}  3\delta_T} \right) }+\log\left(\frac{1}{\delta_T\left(1-e^{f(\theta^*)}\right)}\right)\right)
\end{align}
where $K$ is defined as \eqref{eq:definition_K}.

Let $\delta_T=1/T^2$; the sequence of probability values satisfy $\sum^{\infty}_{T=1}\delta_T<\infty$, so the Borel-Cantelli lemma implies that, almost surely,
    \begin{align}
         \limsup_{T\to\infty}\frac{1}{T} \sum_{t=1}^{T} d(X_t, \hat{X}_t) \leq &D.
    \end{align}
\end{document}